\documentclass[useAMS,usenatbib,usegraphicx]{mn2e}
\usepackage[english]{babel}
\usepackage{multirow}

\newcommand{\OIII}{[O{\sc iii}]}

\newcommand{\NII}{[N{\sc ii}]}
\newcommand{\SII}{[S{\sc ii}]}

\newcommand{\HII}{H{\sc ii}}

\newcommand{\Ha}{H$\alpha$}
\newcommand{\Hb}{H$\beta$}
\newcommand{\OIIIHb}{[O{\sc iii}]5007/H$\beta$}
\newcommand{\NIIHa}{[N{\sc ii}]6548,6584/H$\alpha$}
\newcommand{\SIIHa}{[S{\sc ii}]6717,6731/H$\alpha$}

\newcommand{\myr}{\,\mbox{mas}\,\mbox{yr}^{-1}}
\newcommand{\kms}{\,\mbox{km}\,\mbox{s}^{-1}}

\title
[Cyg OB1. Star cluster vdB~130.] {Star-forming regions at the periphery of
the supershell surrounding the Cyg OB1 association. I. The star cluster vdB~130
and its ambient gas and dust medium}

\author[Sitnik et al.]{
   T.G.~Sitnik$^{1}$\thanks{E-mail: sitnik@sai.msu.ru},
   O.V.~Egorov$^{1}$,
   T.A.~Lozinskaya$^{1}$,
   A.V.~Moiseev$^{1,3}$,
   A.S.~Rastorguev$^{1}$,\newauthor
   A.M.~Tatarnikov$^{1}$,
   A.A.~Tatarnikova$^{1}$,
   D.S.~Wiebe$^{2}$\thanks{E-mail: dwiebe@inasan.ru},
   M.V.~Zabolotskikh$^{1}$ \\
 $^{1}$Lomonosov Moscow State University, Sternberg Astronomical Institute,
        Universitetsky pr. 13, Moscow 119992, Russia \\
 $^{2}$ Institute of Astronomy (INASAN), Russian Academy of Sciences, Pyatnitskaya str. 48, Moscow 119017, Russia \\
 $^{3}$ Special Astrophysical Observatory, Russian Academy of Sciences, Nizhnij Arkhyz 369167, Russia }

\begin{document}

\date{Accepted 2015 Month 00. Received 2015 Month 00; in original
form 2015 Month 00}

\pagerange{\pageref{firstpage}--\pageref{lastpage}} \pubyear{2015}

\maketitle

\label{firstpage}

\begin{abstract}

Stellar population and the interstellar gas-dust medium in the vicinity of the open star cluster vdB~130 are analysed using optical observations taken with the 6-m telescope of the SAO RAS and the 125-cm telescope of the SAI MSU along with the data of \textit{Spitzer} and \textit{Herschel}. Based on proper motions and \textit{BV} and \textit{JHKs} 2MASS photometric data, we select additional 36 stars as probable cluster members. Some stars in vdB~130 are classified as B stars. Our estimates of minimum colour excess, apparent distance modulus and the distance are consistent with young age (from 5 to 10 Myrs) of the cluster vdB~130. We suppose the large deviations from the conventional extinction law in the cluster direction, with $R_V \sim 4 - 5$. The cluster vdB~130 appears to be physically related to the supershell around Cyg~OB1, a cometary CO cloud, ionized gas, and regions of infrared emission.
There are a few regions of bright mid-infrared emission in the vicinity of vdB~130. The largest of them is also visible on H$\alpha$ and \SII\ emission maps. We suggest that the infrared blobs that coincide in projection with the head of the molecular cloud are \HII\ regions, excited by the cluster B-stars. Some signatures of a shock front are identified between these IR-bright regions.

\end{abstract}


\section{Introduction}

Star formation picture drawn by modern observations is much more intricate
than it used to be a few decades ago. Our current view of the early phases of
pre-stellar and stellar evolution has been expanded significantly by
\emph{Spitzer}, \emph{Herschel}, \emph{HST}, \emph{Chandra}, \emph{XMM}  and
other space-based and ground-based facilities.

Now we have a much better understanding of sequential star formation triggered by
turbulence, feedback from outflows, supernova explosions and expanding \HII\ regions. These and
other factors work together sculpting a fine structure of molecular clouds
with filaments, pillars, blobs, peppered with young stellar objects
(YSOs). Given this complexity, along with large-scale studies, a detailed look at
some specific objects and regions can be useful for distinguishing between
general and particular features of the star formation process.

\begin{figure*}
\includegraphics[width=\linewidth]{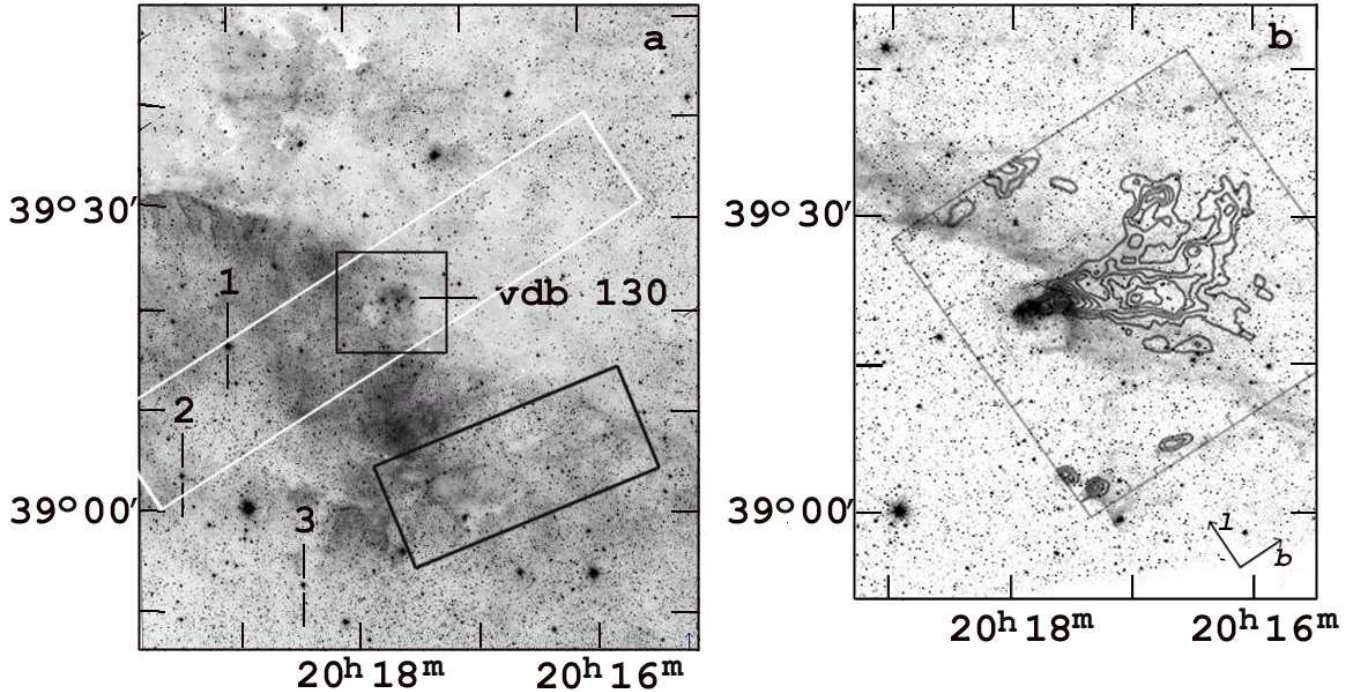}
\caption{(a) A north-western part of the supershell surrounding the Cyg OB1
association as it appears in the optical on the E-DSS2 map. Black rectangles
show the cluster region (top) and the region studied in the work of~\protect\cite{ark13} (bottom).
A white rectangle indicates the region for which
the PV diagram is constructed (see Section~4.1). Numbers mark Cyg OB1 stars
with strong stellar winds: O7e HIP100173 (1), O8 HD193595 (2), and Of HD
228841 (3). (b) The same region
as seen in the infrared on the $3.6\,\mu$m {\em Spitzer} image. Overlaid are
contours of $^{13}$CO J=1-0 emission. The box limits the
area with CO-isocontours from \protect\cite{sch07}.}
\label{fig:Cygpart}
\end{figure*}

 In this paper we present an investigation of the stellar population and gas-dust medium in
the region
located in the north-west border of a supershell surrounding the Cyg OB1
association (Fig.~\ref{fig:Cygpart}a).

The Cyg OB1 supershell with an angular size of $3^\circ \times 4^\circ$ is a part
of Cygnus~X region, the richest and most massive complex of star formation in the
Galaxy. \citet{sch06, sch07, sch11} showed that the Cygnus X South region
contains a large scale network of molecular clouds with ubiquitous
recent and ongoing star formation regions. The area includes several OB
associations and stellar clusters. Their energy injection from UV radiation,
stellar winds and SN explosions could be a driving source for the supershell
formation and expansion. The cloud complex at the south-west of Cygnus X has an extended
shell-like structure, which coincides well with the infrared, optical, and radio supershell
encircling Cyg OB1 and Cyg OB9 associations. Study of the triggered star formation within walls
of this shell can help to reveal the spatially resolved star formation history in
the entire region. In this series of papers we plan to present results on the stellar content,
young stellar objects, and the diffuse matter in the area, putting them into a general context
of the star formation in giant molecular cloud complexes.

In this work we analyze a small region in the north-western wall of Cyg OB1
supershell. It contains the young open cluster vdB~130 characterized by an
anomalously strong and non-uniform reddening~\citep{rac74,moffat,Led02}, the
reflection nebula vdB~130 of the same name~\citep{van66}, also referred to as
GN 20.16.0 \citep{mag}, multiple compact and extended infrared (IR) sources,
and a molecular cloud of a cometary shape seen in CO emission \citep[cloud A in][]{sch07}.

The goal of this study is to revise the stellar content of
the cluster vdB~130 (in a view of new optical and IR data) and to analyse the
physical relationship between various components of
stellar, gaseous, and dust populations in
the region of sequential star formation.

This paper represents a follow up of our previous studies of triggered star
formation regions in the north-west wall of the Cyg OB1 supershell. In the
work of \cite{ark13} a family of cometary globules with a few patches of star formation
has been investigated in a region, shown with a lower black rectangle in Fig.~\ref{fig:Cygpart}a.

Section~2 describes the spectroscopic, photometric, and interferometric
observations. Section~3 presents the results of our revision of the stellar
composition of vdB~130 based on the data of the 2MASS survey, the UCAC4
catalogue of proper motions, and the spectral types that we determined for a
number of stars. In Sections~4.1 and 4.2 we analyse the structure,
kinematics, and the emission spectrum of the interstellar medium (ISM) in the
vicinity of vdB~130, and possible links between various components of the
stellar population and the ISM in the region. In Section~4.3 we discuss the
nature of the IR radiation of the interstellar medium in the neighbourhood of
vdB~130 based on \emph{Spitzer} and \emph{Herschel} archival data. The
concluding section summarizes the main results of the study.

\section{Observations and data reduction}

This study is based on optical observations performed with the
\hbox{6-m} telescope of Special Astrophysical Observatory of
Russian Academy of Sciences (SAO RAS) and the 125-cm telescope of
the Crimean Laboratory of Sternberg Astronomical Institute,
Lomonosov Moscow State University (SAI MSU) as well as on archival
IR data obtained with \textit{Spitzer} and \textit{Herschel} space
observatories.
The log of our observations is
presented in Table~\ref{tab:obs_data}, where the exposure time
($\mathrm{T_{\rm exp}}$), the field of view (FOV), the final angular
resolution ($\theta$), the final spectral resolution ($\delta\lambda$) and
the bandwidth (FWHM) of the used filter are indicated.

\subsection{Spectral observations}

\begin{table*}
\caption{Summary of observational data}
\label{tab:obs_data}
\begin{tabular}{lcllcclc}
\hline
Data set & Filter or Disperser  & Date of obs  & $\mathrm{T_{exp}}$ (sec)  & FOV  & $\theta$ $('')$ &  Sp. range  & $\delta\lambda$ or FWHM (\AA) \\
\hline
Field \#1 & FN655 & 2014/10/16 & 360  & $6.1'\times6.1'$  &  3.5 &    \Ha$+$\NII\ & 97 \\
                   & FN674 & 2014/10/16 & 360 & $6.1'\times6.1'$ & 3.5 &     \SII\ & 60 \\
                   & FN641 & 2014/10/16 & 160  & $6.1'\times6.1'$  &  3.5 &     continuum & 179 \\
                   & FN712 & 2014/10/16 & 160  & $6.1'\times6.1'$  & 3.5 &     continuum & 209 \\
Field \#2 & FN655 & 2014/10/16 & 180  & $6.1'\times6.1'$  &  4.1 &    \Ha$+$\NII\ & 97 \\
                   & FN674 & 2014/10/16 & 180 & $6.1'\times6.1'$ & 4.1 &     \SII\ & 60 \\
                   & FN641 & 2014/10/16 & 80  & $6.1'\times6.1'$  &  4.1 &     continuum & 179 \\
                   & FN712 & 2014/10/16 & 80  & $6.1'\times6.1'$  & 4.1 &     continuum & 209 \\
long-slit \#1    & VPHG1200@540   & 2014/10/16  & 1800 & $6.1'\times1.0''$  & 2.6  & 3690--7280\AA  & 4.5 \\
long-slit \#2    &  VPHG1200G    & 2014/08/03  & 1800 & $6.1'\times1.0''$ & 1.3  & 3690--5680\AA  & 5.4 \\
long-slit \#3    &  VPHG550G   & 2014/08/22  & 1800  & $6.1'\times0.75''$  & 1.9  & 3710--7890\AA  & 9.5 \\
long-slit \#4    & VPHG1200G    & 2014/08/03  & 1800  & $6.1'\times1.0''$  & 1.4  & 3690--5680\AA  & 5.4 \\
long-slit \#5    &  VPHG1200G    & 2014/07/31  & 870 & $6.1'\times1.0''$   & 1.5  & 3690--5680\AA  & 5.5 \\
FPI                 &            & 1990--2001 & 1800 & $10'\times10'$  & 3--4  & \Ha & 0.2--0.35 \\
  \hline
\end{tabular}
\end{table*}

The spectral observations were performed with the 6-m telescope using the
multi-mode SCORPIO (Spectral Camera with Optical Reducer for
Photometrical and Interferometrical Observations) focal reducer
\citep{scorpio} and its new SCORPIO-2 version \citep{scorpio2} operating in
the long-slit spectrograph mode. The scale along the slit was 0.36~arcsec per
pixel. The CCD detectors \hbox{EEV 42-40} in SCORPIO and \hbox{E2V 42-90}
(with the low sensitivity at the blue end of the spectrum) in SCORPIO-2 were
employed. We observed the region around vdB~130 with five slit positions. The
information on all the obtained spectra is summarized
in~Table~\ref{tab:obs_data}.

Data reduction was performed in a standard way using the \textsc{idl}
software package developed at the SAO RAS for reducing long-slit
spectroscopic data obtained with SCORPIO, which includes bias subtraction,
line curvature and flat field corrections, linearization, and
air-glow lines subtraction. Dispersion correction of the spectra was
performed using the reference spectrum of a He-Ne-Ar lamp obtained during the
observations.

The observed region is densely filled with nebular emission (see
Section~\ref{sec:ism}). Because of that we were forced to
construct the night-sky spectrum model using only a few positions
along the slit in the direction of the dark cloud where the
contribution of nebular emission to the spectrum is less
significant. We interpolated these values of air-glow line
intensities for all positions along the slit. The slit \#3 does
not cross the dark cloud, so in order to subtract the air-glow
lines, we observed a nearby blank field where the nebular emission
is absent. To convert the spectra to the absolute intensity scale,
we observed the spectrophotometric standards BD+25d4655,
BD+28d4211 and BD+33d2642 immediately after the object at a close
zenith distance. Gaussian fitting was applied to measure the
integrated fluxes of emission lines. For this purpose we
used our own \textsc{idl} software adopted for SCORPIO data. In
order to estimate uncertainties of the gaussian approximation a
number of synthetic spectra with predefined signal-to-noise ratios
were generated and fitted with the \textsc{idl} procedure. Then,
the obtained uncertainty for a particular signal-to-noise ratio
was used as an estimate of the uncertainty of the method.

To estimate the \HII\ line-of-sight velocities in the region, we used the results of
long-term (1990--2001) H$\alpha$ observations of the Cygnus gas-dust complex
performed with the Fabry--Perot interferometer (FPI) mounted in the Cassegrain focus
of the 125-cm reflector of the Crimean Laboratory of SAI MSU
\citep{loz97,loz98}. The parameters of the instruments are listed in
Table~\ref{tab:obs_data}. We fitted the line profiles by one or several
Gaussian curves assuming that the halfwidth of each profile component should be
greater than that of the instrumental contour and signal-to-noise ratio is no less than 5.
The peak velocities of each profile component were determined.

\subsection{Narrow-band images}

Optical images of two overlapping fields around vdB~130 were taken at the
primary focus of the 6-m telescope of SAO RAS with SCORPIO-2 multi-mode focal
reducer \citep{scorpio2} using filters FN655 and FN674, corresponding to \Ha\
and \SII\ emission lines with central wavelengths of 6559 and 6733 \AA\, and
FWHM = 97 and 60 \AA, respectively. The resolution was 0.36 arcsec per pixel. Since the FWHM of the used FN655 filter is broader than the
separation between \Ha\ and \NII\ emission lines, the image in this filter is
contaminated by \NII\ 6548, 6584~\AA\ emission. Given the results of our
spectral observations (see Section~\ref{sec:ism}), and taking into account
the FN655 filter transmission, we conclude that the contribution of \NII\
lines to the obtained image may be up to 0.4 of the pure \Ha\ emission.
Hereafter we indicate with \Ha\ the emission in the \Ha\ line with this additional
contribution due to the [NII] lines.

We used broader FN641 and FN712 filters centered on the continuum near the
\Ha\ and \SII\ emission lines to subtract the stellar contamination from the
images obtained during the same night. In order to calibrate the emission
line images to energy fluxes we observed the standard star BD+25d4655
immediately after observing the cluster region.

After the data reduction has been made in \textsc{idl} in the
standard way (bias subtraction, flat field normalization, cosmic
ray rejection, background estimation and subtraction) we combined
images of both observed fields around vdB~130 in mosaics in \SII\
and \Ha\ lines using our own \textsc{idl} procedure and
assuming positions of several stars in one frame as a reference.

\subsection{Archival data}

This work is partially based on  data from
\textit{Spitzer} and \textit{Herschel} space telescopes. The region studied
was observed by \textit{Spitzer} space telescope within the framework of the
`A~Spitzer Legacy Survey of the Cygnus-X Complex' program \citep{cygX}. The
images at $3.6\,\mu$m, $4.5\,\mu$m, $5.8\,\mu$m, $8.0\,\mu$m, and $24\,\mu$m
were downloaded from the project
web-page\footnote{https://www.cfa.harvard.edu/cygnusX/} at the
NASA/IPAC Infrared Science Archive. The \textit{Herschel} archival data
obtained within the framework of the `Hi-GAL: The Herschel Infrared Galactic
Plane Survey'\footnote{https://hi-gal.iaps.inaf.it/higal/}
program~\citep{Molinari10} were downloaded from the Herschel Science
Archive\footnote{http://www.cosmos.esa.int/web/herschel/science-archive}.

In this study we also used images from the 2MASS IR survey \citep{skr06} and
stellar proper motions  data from the UCAC4 catalogue \citep{ucac4}.

\section{Open star cluster vdB~130}

\begin{figure}
\includegraphics[width=\linewidth]{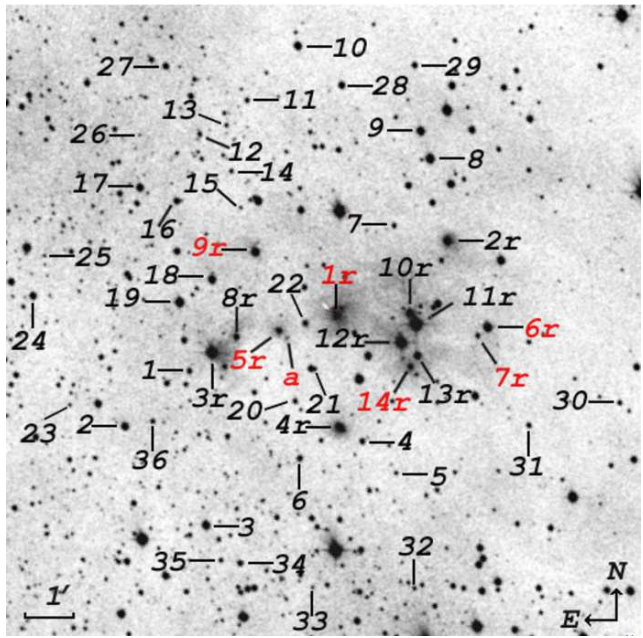}
\caption{Star cluster vdB~130. Stars from the original \citet
{rac74} paper are marked with indices from 1r to 14r. Stars
selected in this study based on their proper motions are marked
with numbers from 1 to 36. Stars from \citet{rac74} list with
proper motions outside of the adopted intervals are shown with red
colour. The letter \emph{a} marks the star K4 projected onto an
extended IR source (Blob E; see section 4.3). The central star
BD+38 3993 of the reflection nebula is marked as 11r.}
\label{fig:vdB130}
\end{figure}

Initially designation `vdB~130' has been applied to a reflection
nebula associated with the BD+38 3993 star (HD 228789, vdB~130a)
of spectral class B1-B2 \citep{van66, rac68}. The reflection
nebula vdB~130 can be seen on the optical maps shown in Figures
\ref{fig:Cygpart}a and~\ref{fig:vdB130}. \citet{rac74} used
photometric and spectroscopic data to select the group of 14 stars
within $6^{\prime}$ from the nominal cluster centre, which could
be considered as possible members of the sparse embedded cluster
vdB~130 (Fig.~\ref{fig:vdB130}). He noted an anomalous extinction
law with ${A_{V}}/{E(B-V)} = 8.1 \pm 1.2$, probably due to
inhomogeneity of the interstellar medium in the proximity of the
cluster. \citet{khar13}, with pipe-line processing technique,
estimated the cluster parameters using data from PPMXL catalogue
\citep{ppmxl} and considering stars with proper motion components
close to the centroid with $\langle\mu_{\alpha},
\mu_{\delta}\rangle \approx (-2.22, -4.39) \myr$ as cluster
members. From 2MASS JHKs photometric data they derived a cluster
age $\log(t) \approx 7.4$, distance $d \approx 1.6$ kpc, colour
excess $E(J-H) \approx 0.29$ mag, infrared interstellar absorption
$A_{Ks} \approx 0.30$ mag, and apparent radius of about
$6^{\prime}$.

\subsection{Revision of the stellar content of vdB~130}

Most possible members of vdB~130 move parallel to the galactic
plane with mean proper motions $\langle\mu_l, \mu_b\rangle \approx
(-4.9, -0.6) \myr$ \citep{khar13}. In order to estimate the
contribution of the disk differential rotation and that of the Sun
peculiar velocity to the observed proper motions of cluster stars
for heliocentric distances $r_{hel} \sim 1.5 - 2.5$~kpc, the
parameters of Milky Way disk rotation and Sun's velocity
components with respect to the LSR (Local Standard of Rest) were
taken from a detailed kinematical analysis of young stellar
populations performed by \citet{zabol02} and \citet{MD2009}. If we
use a currently adopted value of the Sun's distance from the Milky
Way centre, $\approx 8.3$~kpc, all fields located along the line
of sight in Cygnus directions with heliocentric distances in the
$r_{hel} \approx 1.5 - 2.5$~kpc range, are nearly equally distant
from the galactic centre, $\approx 8.1$~kpc. From the 3D
Bottlinger equations we estimate mean apparent proper motion at
1.5--1.8~kpc heliocentric distances as $\langle\mu_l(rot), \;
\mu_b(rot)\rangle \approx  (-5.4, -0.9) \myr$.

\begin{figure}
\includegraphics[width=\linewidth]{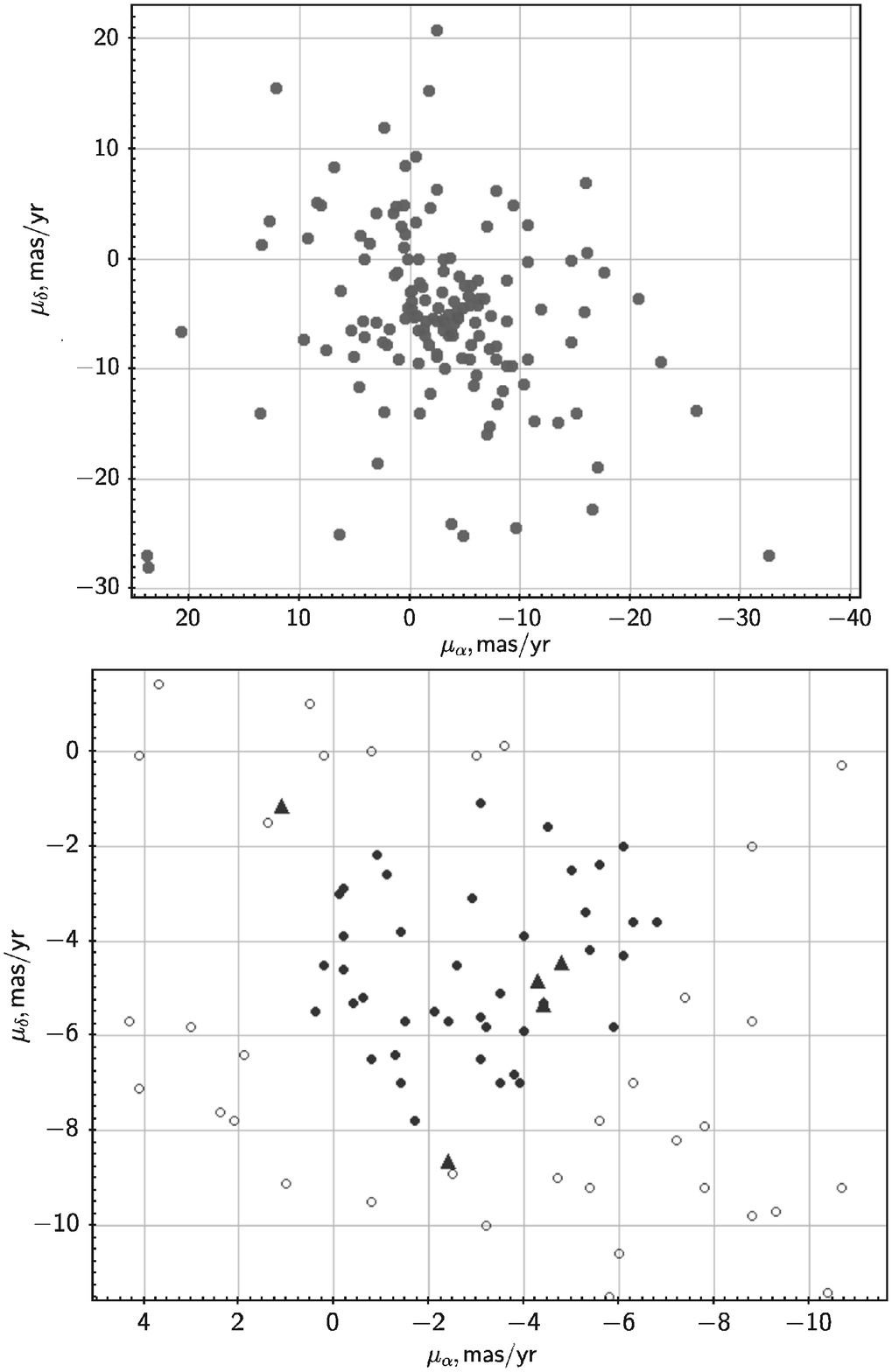}
\caption{(Top) Proper motion ($\mu_\alpha, \; \mu_\delta$) diagram for stars
from the UCAC4 catalogue in the circle within $6^{\prime}$ from the centre of
the field. (Bottom) Central part of the diagram. Filled circles indicate
possible cluster members selected in this study. Shown with triangles are
stars with the most accurate proper motions taken from the TYCHO2 catalogue.
Field stars are shown with open circles.} \label{fig:large}
\end{figure}

These values are in a very good agreement with the proper motion
components for vdB~130 derived by \citet{khar13}.
Therefore, we cannot exclude that stars assigned to the
open cluster by \citet{khar13} represent in fact a non-bound
stellar population of the extended association Cyg OB1,
participating in overall galactic differential rotation. Moreover,
the accuracy of available proper motions in this field does not
allow distinguishing reliably true members of an unbound OB
association from foreground and background field stars.

\begin{table*}
\caption{Candidate members of vdB~130}\label{tab:cluster}
\begin{scriptsize}
\begin{tabular}{l|l|l|l|l|l|l|l|l|l|l}

\hline
Star \# & \,\,\,\,UCAC4     &\,\,\,\,\,\,\,\,\,\,\,\,\,\,$\mu_\alpha$ &\,\,\,\,\,\,\,\,\,\,\,\,\,\,$\mu_\delta$ &\,\,\,\,\,\,Tycho-2     &\,\,\,\,\,\,2Mkey    &\,\,\,\,\,\,\,\,\,\,\,\,\,$J$ (mag)  &$H$ (mag)  &$K$ (mag)  &$B$ (mag)  &$V$ (mag)  \\
\hline
 1 & 647-086881 & $-1.5 \pm 6.0$    & $-5.7 \pm 5.9 $  &             & 305146782 &$13.2\,\,\,\,\,\, \pm 0.03 $ & $12.856 \pm 0.03$  & $12.666 \pm 0.03$  & 16.031 & 15.332\\
 2 & 647-086916 & $-5.4 \pm 2.2$    & $-4.2 \pm 2.2 $  &             & 305146998 &$12.01\,\,\, \pm  0.02 $ & $11.679 \pm 0.02$  & $11.56\,\,\, \pm 0.02$  & 14.212 & 13.582\\
 2r& 647-086761 & $- 5.0 \pm 2.2$   & $-2.5 \pm 4.5 $  &             & 305106035 &$11.703 \pm 0.02 $ & $11.352 \pm 0.02$  & $11.225 \pm 0.02 $ & 13.907 & 13.28 \\
 3 & 647-086873 & $- 4.5 \pm 1.9$   & $-1.6 \pm 2.4 $  &             & 305105199 &$12.045 \pm 0.02 $ & $11.95\,\,\, \pm 0.02$  & $11.853 \pm 0.02 $ & 13.555 & 13.067\\
 3r& 647-086871 & $- 4.8 \pm 0.9$   & $-4.5 \pm 0.6 $  &3151-00092-1 & 305105710 &$\,\,\,9.591 \pm 0.02 $ &  $\,\,\,9.421 \pm 0.02$  & $ \,\,\,9.33\,\,\, \pm 0.02$  & 11.597 & 10.99 \\
 4 & 647-086802 & $- 3.5 \pm 6.0$   & $-5.1 \pm 2.8 $  &             & 305105443 &$12.615 \pm 0.02 $ & $12.33\,\,\, \pm 0.02$  & $12.203 \pm 0.02  $& 15.636 & 14.883\\
 4r& 647-086813 & $- 4.4 \pm 1.9$   & $-5.3 \pm 3.7 $  &             & 305105485 &$11.376 \pm 0.02 $ & $11.086 \pm 0.02$  & $10.909 \pm 0.02$  & 13.852 & 13.125\\
 5 & 647-086786 & $- 0.2 \pm 4.5$   & $-3.9 \pm 4.3 $  &             & 305105367 &$13.047 \pm 0.02 $ & $12.316 \pm 0.02$  & $12.164 \pm 0.02$  &       &       \\
 6 & 647-086831 & $- 3.2 \pm 4.8$   & $-5.8 \pm 4.5 $  &             & 305105399 &$12.948 \pm 0.02 $ & $12.618 \pm 0.02$  & $12.469 \pm 0.03 $ & 15.634 & 14.914\\
 7 & 647-086791 & $- 2.9 \pm 5.4$   & $-3.1 \pm 4.8 $  &             & 305106069 &$13.761 \pm 0.03 $ & $13.408 \pm 0.02$  & $13.282 \pm 0.04 $ & 16.781 & 15.847\\
 8 & 648-090623 & $- 3.1 \pm 2.1$   & $-1.1 \pm 1.7 $  &             & 305106265 &$11.65\,\,\, \pm 0.04 $ & $11.382 \pm 0.02$  & $11.32\,\,\, \pm 0.02 $ & 13.295 & 12.727\\
 8r& 647-086860 & $- 6.1 \pm 4.2$   & $-4.3 \pm 4.2 $ &             & 305105753 &$11.994 \pm 0.02 $ & $11.4\,\,\,\,\,\, \pm 0.03 $ & $10.962 \pm 0.02 $ & 15.466 & 14.642\\
 9 & 648-090627 & $- 0.6 \pm 1.6$   & $-5.2 \pm 2.7 $  &             & 305106343 &$11.635 \pm 0.02 $ & $11.311 \pm 0.02$  & $11.239 \pm 0.02$  & 13.801 & 13.09 \\
10 & 648-090649 & $- 4.0 \pm 2.2$   & $-5.9 \pm 2.0 $  &             & 305106565 &$12.028 \pm 0.02 $ & $11.75\,\,\, \pm 0.02 $ & $11.676 \pm 0.02 $ & 13.986 & 13.352\\
10r& 647-086781 & $- 4.0 \pm 3.3$   & $-3.9 \pm 4.0 $  &             & 305105837 &$11.176 \pm 0.02 $ & $10.808 \pm 0.02$  & $10.652 \pm 0.02 $ &       &       \\
11 & 648-090655 & $- 3.8 \pm 4.6$   & $-6.8 \pm 4.4 $  &             & 305106417 &$13.168 \pm 0.02 $ & $12.792 \pm 0.02$  & $12.572 \pm 0.03 $ & 16.971 & 15.835\\
11r& 647-086776 & $- 4.3 \pm 0.8$   & $-4.9 \pm 0.5 $  &3151-00165-1 & 305105806 &$\,\,\,8.98\,\,\,\pm 0.03 $ &  $\,\,\,8.816 \pm 0.02 $ &  $\,\,\,8.745 \pm 0.02 $ & 10.755 & 10.214\\
12 & 648-090662 & $- 0.4 \pm 5.9$   & $-5.3 \pm 6.0 $  &             & 305145860 &$12.141 \pm 0.02 $ & $11.543 \pm 0.02$  & $11.354 \pm 0.02 $ & 16.503 & 15.197\\
12r& 647-086785 & $- 4.4 \pm 0.9$   & $-5.4 \pm 1.1 $  &3151-00109-1 & 305105748 &$\,\,\,9.79\,\,\, \pm 0.02 $ &  $\,\,\,9.621 \pm 0.02 $ & $ \,\,\,9.526 \pm 0.02 $ & 11.54 & 11.022\\
13 & 648-090657 & $- 6.8 \pm 5.2$   & $-3.6 \pm 5.8 $  &             & 305106346 &$14.05\,\,\, \pm 0.02 $ & $13.73\,\,\, \pm 0.04 $ & $13.534 \pm 0.05 $ &       &       \\
13r& 647-086775 & $- 3.9 \pm 4.4$   & $-7.0 \pm 7.0 $  &             & 305105714 &$11.738 \pm 0.02 $ & $11.379 \pm 0.02$  & $11.236 \pm 0.02 $ &       &       \\
14 & 647-086865 & $- 6.1 \pm 4.4$   & $-2.0 \pm 4.5 $  &             & 305106215 &$13.2\,\,\,\,\,\, \pm 0.02 $ & $12.664 \pm 0.02$  & $12.457 \pm 0.03 $ & 17.236 & 15.873\\
15 & 647-086859 & $- 1.3 \pm 5.5$   & $-6.4 \pm 5.3 $  &             & 305106104 &$14.472 \pm 0.04 $ & $13.829 \pm 0.04 $ & $13.734 \pm 0.05 $ &       &       \\
16 & 647-086890 & $- 2.1 \pm 2.7$   & $-5.5 \pm 3.1 $  &             & 305146111 &$13.146 \pm 0.03 $ & $12.919 \pm 0.03 $ & $12.786 \pm 0.03 $ & 15.409 & 14.717\\
17 & 647-086910 & $- 2.6 \pm 2.3$   & $-4.5 \pm 3.7 $  &             & 305146062 &$12.202 \pm 0.02 $ & $11.907 \pm 0.02 $ & $11.829 \pm 0.02 $ & 14.278 & 13.713\\
18 & 647-086874 & $- 1.4 \pm 2.6$   & $-3.8 \pm 2.4  $ &             & 305105912 &$11.632 \pm 0.02 $ & $11.385 \pm 0.02 $ & $11.233 \pm 0.02 $ & 13.898 & 13.214\\
19 & 647-086887 & $- 6.3 \pm 0.9$   & $-3.6 \pm 1.1 $  &             & 305146507 &$11.091 \pm 0.02 $ & $10.943 \pm 0.02 $ & $10.839 \pm 0.02 $ & 13.003 & 12.406\\
20 & 647-086833 & $-0.2 \pm 4.6$   & $-4.6 \pm 4.4 $  &             & 305105567 &$13.154 \pm 0.03 $ & $12.775 \pm 0.03 $ & $12.58\,\,\, \pm 0.03$  & 16.57 & 15.55 \\
21 & 647-086826 & $\,\,\,\,0.4 \pm 3.5$     & $-5.5 \pm 2.4 $  &             & 305105664 &$12.03\,\,\, \pm 0.03 $  & $11.581 \pm 0.04 $ & $11.402 \pm 0.03 $ & 15.08 & 14.146\\
22 & 647-086829 & $- 0.8 \pm 4.3$   & $-6.5 \pm 4.3 $  &             & 305105800 &$12.741 \pm 0.99 $ & $12.166 \pm 0.03 $ & $11.925 \pm 0.02 $ & 16.35 & 15.262\\
23 & 647-086940 & $- 2.4 \pm 6.3 $  & $-5.7 \pm 6.1 $  &             & 305146927 &$12.128 \pm 0.02 $ & $11.252 \pm 0.02 $ & $10.956 \pm 0.02 $ &       &       \\
24 & 647-086954 & $- 5.6 \pm 4.9 $  & $-2.4 \pm 1.8 $  &             & 305146496 &$12.121 \pm 0.02 $ & $11.708 \pm 0.02 $ & $11.457 \pm 0.02 $ & 14.536 & 13.87 \\
25 & 647-086949 & $- 3.1 \pm 6.5 $  & $-6.5 \pm 6.2  $ &             & 305146341 &$12.857 \pm 0.03 $ & $12.081 \pm 0.03 $ & $11.835 \pm 0.03 $ &       &       \\
26 & 648-090681 & $- 0.9 \pm 7.2$   & $-2.2 \pm 6.8  $ &             & 305145870 &$12.397 \pm 0.02 $ & $11.467 \pm 0.02 $ & $11.183 \pm 0.02 $ &       &       \\
27 & 648-090678 & $\,\,\,\,0.2 \pm 3.9$     & $ -4.5 \pm 4.8 $  &             & 305145626 &$13.078 \pm 0.02 $ & $12.736 \pm 0.02$  & $12.655 \pm 0.03$  & 15.483 & 14.753\\
28 & 648-090639 & $- 5.3 \pm 2.2 $  & $-3.4 \pm 2.4 $  &             & 305106466 &$12.516 \pm 0.02 $ & $12.141 \pm 0.02 $ & $12.072 \pm 0.02 $ & 14.73 & 13.959\\
29 & 648-090628 & $- 3.1 \pm 4.7 $  & $-5.6 \pm 4.4 $  &             & 305106531 &$13.004 \pm 0.02 $ & $12.611 \pm 0.02 $ & $12.463 \pm 0.02 $ & 15.776 & 14.911\\
30 & 647-086696 & $- 1.1 \pm 4.5 $  & $-2.6 \pm 4.4 $  &             & 305036100 &$13.573 \pm 0.02 $ & $13.161 \pm 0.02 $ & $13.095 \pm 0.03 $ & 16.62 & 15.574\\
31 & 647-086727 & $- 0.1 \pm 4.6 $  & $-3.0 \pm 4.6 $  &             & 305105507 &$13.653 \pm 0.03 $ & $13.263 \pm 0.04 $ & $13.203 \pm 0.04 $ & 16.397 & 15.427\\
32 & 647-086773 & $- 5.9 \pm 4.3 $  & $-5.8 \pm 4.3 $  &             & 305105002 &$12.236 \pm 0.02 $ & $11.49\,\,\, \pm 0.02$  & $11.275 \pm 0.02$  &       &       \\
33 & 647-086823 & $- 1.4 \pm 4.8 $  & $-7.0 \pm 4.6 $  &             & 305104995 &$12.421 \pm 0.02 $ & $11.567 \pm 0.02 $ & $11.295 \pm 0.02 $ &       &       \\
34 & 647-086856 & $- 3.5 \pm 4.0 $  & $-7.0 \pm 1.3 $  &             & 305105078 &$12.536 \pm 0.02 $ & $12.221 \pm 0.02 $ & $12.074 \pm 0.02 $ & 15.613 & 14.701\\
35 & 647-086866 & $- 0.2 \pm 5.9 $  & $-2.9 \pm 6.3 $  &             & 305105086 &$13.673 \pm 0.03 $ & $13.354 \pm 0.03 $ & $13.144 \pm 0.04 $ & 17.342 & 15.927\\
36 & 647-086897 & $- 1.7 \pm 5.8 $  & $-7.8 \pm 6.1$   &             & 305146976 &$13.636 \pm 0.03 $ & $13.261 \pm 0.03 $ & $13.086 \pm 0.04 $ & 16.644 & 15.926\\
\hline 1r & 647-086816 & $ 23.9 \pm 18.8$  & $-27.0 \pm 9.0 $   &
& 305105825
&$11.594 \pm 0.03 $ & $10.744 \pm 0.02 $ & $10.401 \pm 0.02 $ & 15.085 & 14.225\\
5r & 647-086841 & $- 7.0 \pm 4.4  $  & $ 2.9 \pm 4.4$   &             &
305105774
&$11.988 \pm 0.04 $ & $11.435 \pm 0.04 $ & $11.108 \pm 0.04 $ & 15.744 & 14.741\\
6r & 647-086746 & $ 53.1 \pm 2.8 $  & $34.5 \pm 1.8$   &             &
305105807
&$10.211 \pm 0.02 $ &  $9.666 \pm 0.02 $ &  $9.56 \pm 0.02 $  & 13.152 & 12.124\\
7r & 647-086750 & $ 32.4 \pm 5.7 $  & $-8.5 \pm 6.1$   &             &
305105772
&$13.432 \pm 0.02 $ &  $12.924 \pm 0.03$ & $12.835 \pm 0.03$  &      &         \\
9r & 647-086849 & $- 4.7 \pm 1.7 $  & $- 9.0 \pm 3.3$   &             &
305105989
&$11.330 \pm 0.02 $ &  $11.092 \pm 0.02$ & $11.025 \pm 0.02$  & 13.383 & 12.752\\
14r& 647-086779 & $              $  & $            $   &             &
305105685
&$12.536 \pm 0.02 $ & $12.163 \pm 0.02 $ & $11.945 \pm 0.02$  &      &         \\

\hline
\end{tabular}

\end{scriptsize}
\end{table*}

\citet{Led02} also used 2MASS data to study the stellar population and the
structure of vdB~130, along with other 21 clusters in the Cygnus region.
Unfortunately, the lack of detailed data on this cluster does not allow
estimating its parameters reliably. It should be noted that application of
King's empirical density law \citep{king62} to describe the apparent
distribution of stars in dense and highly non-uniform stellar fields
distorted by inhomogeneous absorption is unlikely able to provide reliable
estimates of important structure cluster parameters. \citet{Led02}
used the cluster distance derived by \citet{rac74} and claimed vdB~130 as the
richest cluster among those studied by them in the Cygnus region. It should
be noted, however, that they used no kinematical criteria to select probable
cluster members.

Despite the insufficient evidence supporting the nature of vdB~130 as a true
gravitationally bound stellar cluster, further study of the stellar content
of this star field, located at the distance of about 30--40 parsecs from the
centre of the Cyg~OB1 association, seems to be worthwhile. To find
new possible members of vdB~130 and confirm the membership of stars
earlier attributed to this cluster by \citet{rac74} we reconsidered the data
of 175 stars in a rather extended region with the diameter of 12~arcmin and
central coordinates
$\alpha_{2000}\sim20^{\mathrm{h}}17^{\mathrm{m}}50^{\mathrm{s}}$,
$\delta_{2000}\sim39^{\circ} 21^{\prime}25^{\prime\prime}$, for which stellar
proper motions can be taken from UCAC4 catalogue \citep{ucac4}. The
concentration of stars around the centroid defined by \citet{khar13} is
evident in the top panel of Fig.~\ref{fig:large}. This can be readily
explained by the dominant contribution of disk differential rotation and
solar motion to the apex.

\begin{figure}
\includegraphics[width=\linewidth]{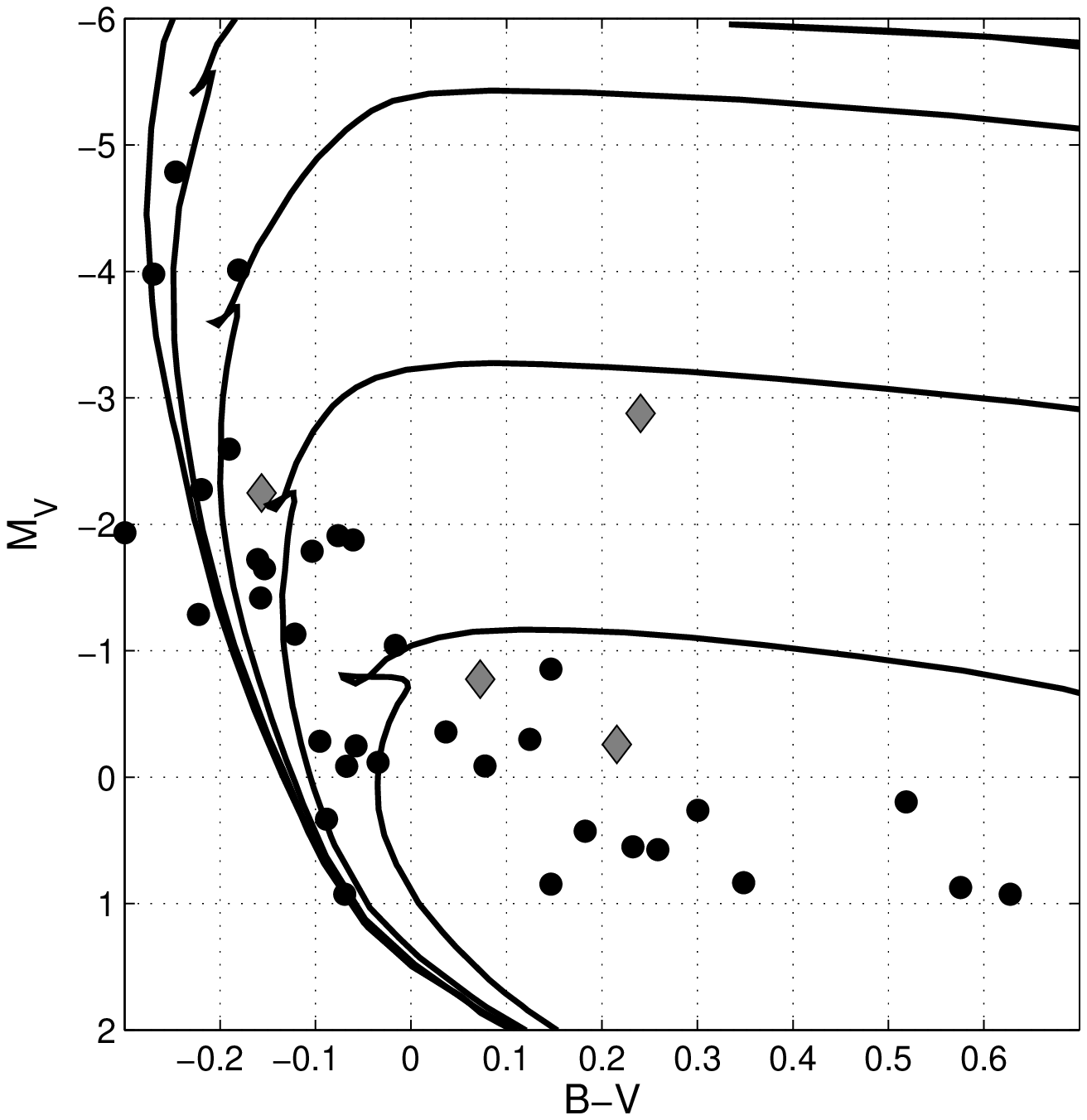}

\includegraphics[width=\linewidth]{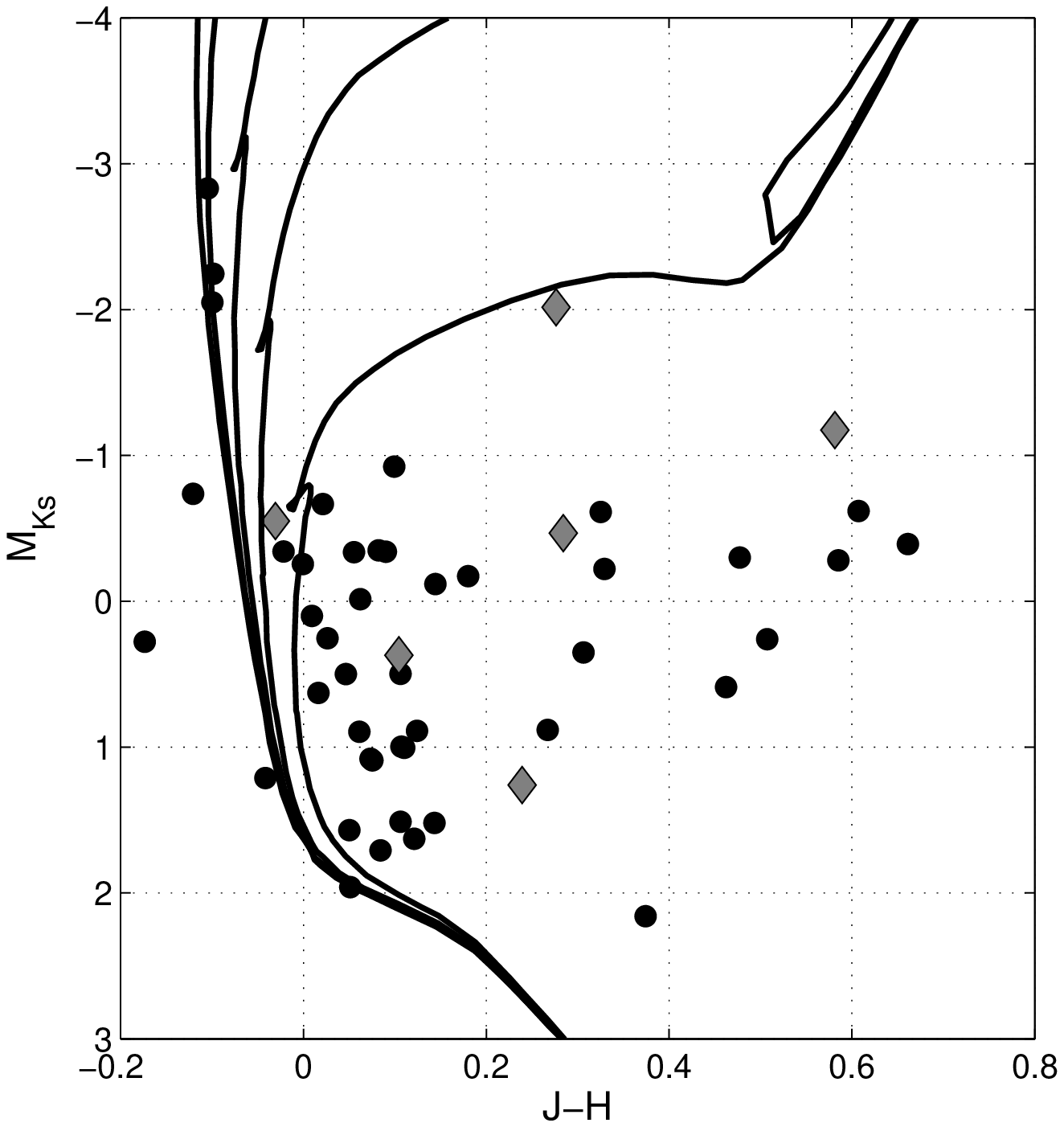}
\caption{Colour-absolute magnitude diagrams $M_V$ vs. $(B-V)$
(top) and $M_{Ks}$ vs. $(J-H)$ (bottom) for vdB~130 possible
members selected by proper motions (circles). Theoretical
isochrones with ages $t = 5, 10, 30, 100, 300$ Myrs (from top to
bottom) are also shown for $BV$ data \citep{padova} and 2MASS
$JHK_s$ data \citep{bbg04}. Diamonds indicate stars from \citet
{rac74} list with large differences in proper motions from the
centroid.} \label{fig:izohrone}
\end{figure}

We selected 36 UCAC4 stars as possible members of the cluster, with proper
motion vectors that differ from the considered centroid by less than $4 \myr$
(see bottom panel of Fig. \ref{fig:large}). This choice of a limiting
value corresponds to the typical proper motion error in the UCAC4 catalogue,
while in most cases the errors range from 1 to 10 mas~yr$^{-1}$. Adopting a larger
threshold of proper motions, we risk of contaminating our sample with a large
number of foreground and background field stars with large transversal
velocities. (Note that the difference of $4 \myr$ at the distance of 1.5~kpc
is equivalent to $28 \kms$ difference in transversal velocity. This value
greatly exceeds the typical velocity dispersion in OB associations.)

Eight of 14 stars from \citet{rac74} original list satisfy our
proper motions criteria (Fig. \ref {fig:vdB130}). The extended
list of possible cluster members is given in
Table~\ref{tab:cluster}. Star identifications, proper
motions and respective errors, 2MASS identifications and JHKs
magnitudes, BV magnitudes (for some stars) were taken from the
UCAC4 catalogue and are printed keeping original catalogue
format. The columns of Table~\ref{tab:cluster} contain the star
number according to \citet{rac74} list (with a letter~`r') and/or
our list, its identification in the UCAC4 catalogue, proper motion
components $\mu_\alpha, \; \mu_\delta$ and corresponding errors,
TYCHO2 catalogue \citep{tycho2} identification (for some stars),
2MASS catalogue identification (2Mkey) and $JHK_s$ magnitudes and
errors, and $BV$ magnitudes. In the bottom part of the table we
show the data for stars from \citet{rac74} list with motions
outside of the adopted intervals (1r, 5r, 6r, 7r, 9r and 14r).
Stars 1r and 5r from \citet{rac74} original list, which are
formally not considered as the cluster members, apparently have
large velocities and move toward each other, with velocity
components relative to vdB~130  $(V_{l}, V_{b}) \approx (-31 \pm
137, -260 \pm 83)$ and ($26 \pm 47, 61 \pm 9$)~$\kms$,
respectively. However, their proper motions are measured with very
large errors, partly due to their asymmetric shapes in the optical
images. Therefore, we cannot decisively exclude these two bright
stars from the vdB~130 cluster members.

Based on the sample of stars shown in Table~\ref{tab:cluster}, we
estimate basic characteristics (colour excess and distance)
vdB~130, using $JHK_s$ and $BV$ magnitudes and isochrone fitting
technique. A colour-absolute magnitude diagram (CMD) $V$ vs. $B-V$
for 34~probable vdB~130 members with available $BV$ data and
Padova isochrones \citep{padova} calculated for solar abundance
and for ages $t = 5, 10, 30, 100, 300$ Myrs are shown on the top
panel of Fig.~\ref{fig:izohrone}. If we take into account
typical photometric uncertainty of $0.03-0.05$~mag and the real
overdensity of bright stars, clearly seen on \citet{khar13} atlas
for vdB~130, the eye-fit gives an estimate for the cluster age of
about 5-10 Myrs, assuming that the best-fit isochrone is a lower
left envelope for stars with lowest colour indices.

Despite a relatively small scatter in proper motions, colour scatter along
the $(B-V)$ axis is quite large (up to 0.6~mag for faint stars), confirming
the substantial differential reddening over the cluster area. Minimal colour
excess $E(B-V) = 0.79 \pm 0.02$~mag, and the apparent distance modulus
$(V-M_V)_{app} = 15.0 \pm 0.3$~mag. Under the assumption of a `normal'
extinction law \citep{car89, mar90}, with $R_{V} = A_{V}/E(B-V) \approx
3.08$, true distance modulus can be estimated as $(V-M_V)_0 \approx 12.6 \pm
0.3$~mag. In this case the distance to vdB~130 stars having minimal reddening
values can be estimated as $3.3 \pm 0.4$~kpc, in excess of \citet{khar13} and
\citet{rac74} estimates. Of course, our estimates of the colour excess and
the distance relate to bluest and brightest stars on the CMDs, which are
expected to lie on the front side of vdB~130. It should also be noted that
\citet{sfd1998} and \citet{sf2011} give total line-of-sight colour excess
$E(B-V)$ values in ranges $4.85-9.6$~mag and $4.2-8.3$~mag, respectively, for
a 5 arcmin radius circle, centred in the field of the vdB 130 cluster. These
data support Racine's conclusion on possible large differential extinction in
this region. 
\begin{figure}
\includegraphics[width=\linewidth]{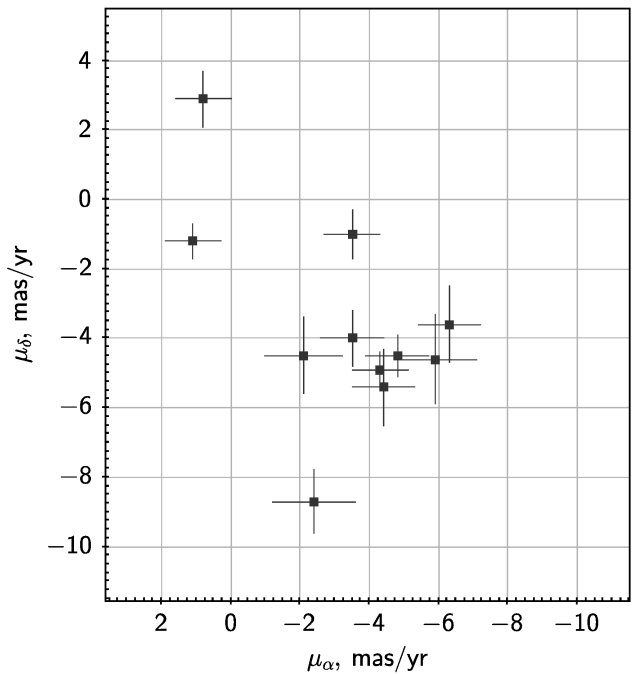}
\caption{A ($\mu_\alpha, \; \mu_\delta$) diagram for stars with most precise
measurements of proper motions.} \label{fig:best}
\end{figure}

A somewhat smaller distance can be derived from infrared data. A
$K_s$ vs. $(J-H)$ colour-magnitude diagram for 36 stars and the
isochrones from \citep{bbg04}, calculated for the same ages as for
optics, are shown on the bottom panel of Fig. \ref{fig:izohrone}.
Using the same fitting technique, we estimate colour excess
$E(J-H) \approx 0.27 \pm 0.02$~mag and apparent distance modulus
$(K-M_{Ks})_{app} \approx 11.57 \pm 0.3$~mag. Under the assumption
of a `normal' extinction law \citep{2Mext,he95}, with $A_{Ks}
\approx 1.13 E(J-H)$, we estimate near infrared (NIR) absorption as
$A_{Ks} \approx 0.31 \pm 0.03$, true distance modulus as
$(K-M_{Ks})_0 \approx 11.26 \pm 0.3$~mag, and the corresponding
cluster distance as $1.8 \pm 0.3$~kpc, in a general agreement with
the result of \citet {khar13}.

The differences between true distance moduli estimated
independently from optical and NIR data can be
attributed to abnormal features of the extinction law in the
cluster direction. Obviously, any deviations from the generally
accepted `normal' extinction law should manifest themselves much
stronger in the optical range than in the infrared. To match the
distance estimates obtained from optical and infrared data, i.e.
from $V$ vs. $(B-V)$ and $K_s$ vs. $(J-H)$ CMDs, we need to assume
that $R_{V}$ in this field differs from the `standard' value of
3.08  and reaches 4.5--4.8, in a good agreement with \citet{moffat}
result. Note, however, that this conclusion is rather preliminary,
because distance moduli in optical and infrared have been really
estimated from only 6--7 stars with the lowest colour excess.
Moreover, some of these stars are located in the upper left part
of the CMD, where the slope of the main sequence and isochrones is
large enough and, thus, vertical shift of isochrones (i.e.
apparent distance modulus) is determined not as reliably as the
colour excess.

We now consider the group of stars with the most accurate
measurements of proper motions. Three stars from TYCHO2
\citep{tycho2} are notable among them, with proper motion errors
smaller than $1.1 \myr$. On the proper motion ($\mu_\alpha,
\mu_\delta$) diagram seven stars form a compact group with $4
\myr$ diameter (Fig. \ref{fig:best}). Four of them, 3r (TYC
3151-00092-1), 11r (TYC 3151-00165-1), 12r (TYC 3151-00109-1), and
19 (2MASS 305146507), are included in our list and they are the
brightest stars in our sample. Mean components of their proper
motion are approximately $\langle\mu_\alpha, \mu_\delta\rangle=
(-4.95, -4.60) \myr$ with the RMS scatter of $(0.9, 0.8) \myr$.
The difference between the mean motion and the \citet{khar13}
centroid (and, at the same time, contribution of disk differential
rotation) is approximately $2.7 \pm 0.9 \myr$. At the cluster
distance ($\approx 1.8$~kpc) this leads to $20 \kms$ difference of
transversal velocities. We can conclude that these brightest
members of the young open cluster vdB~130 associated with the
Cyg~OB1 give an idea on the apparent cluster motion on the sky. To
draw more definite conclusions on the reality of the
gravitationally bound young stellar group additional precise
proper motions measurements and spectroscopic observations are
needed that would allow refining the spectral classification and
provide accurate radial velocity measurements.

\subsection{Spectra of vdB~130 stars}

We used the spectra described in Section 2.1 to determine the
spectral types of the stars 1r, 4r, 5r, 6r, 11r, and 22
(Table~\ref{tab:cluster}) and of the \emph{a} star (Fig.~\ref{fig:vdB130}).
All the stars, except for the last one, have been classified as B
stars. We analysed the \emph{a} star in order to assess its possible
contribution to the emission of the gas-dust nebula studied in Section 4.3 and
classified it as a K4 star. Note that the spectral energy distribution of
this star is practically undistorted by interstellar reddening, which is
unlikely for a star located at the distance of vdB~130 (1.5--2 kpc), and this
further confirms that \emph{a} is a foreground field star.

\begin{figure}
\center{\includegraphics[width=\linewidth]{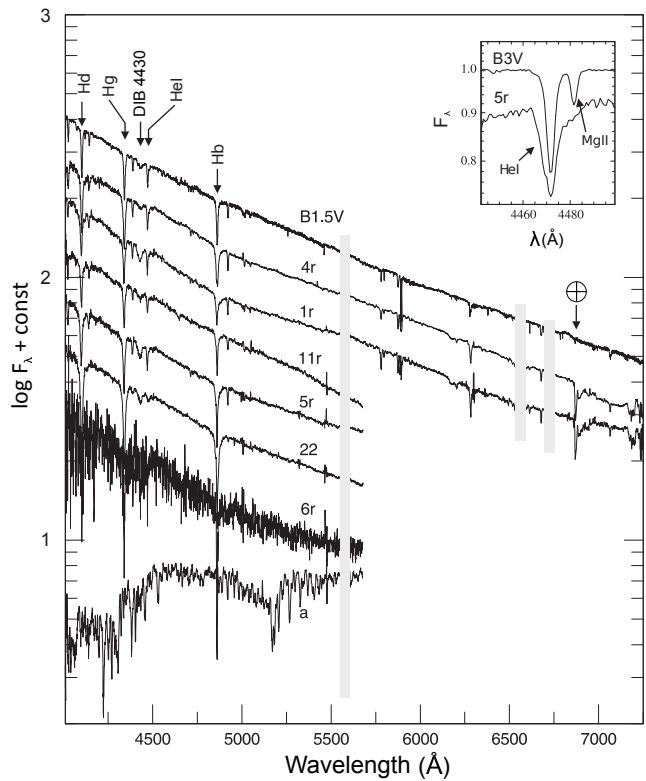}} \caption{Spectra
of some vdB 130 stars corrected for the interstellar reddening and a standard
B1.5V spectrum. Star designations are the same as in Fig.~\ref{fig:vdB130}.
The inset shows an expanded view of  a small part of the star 5r  spectrum
and the standard B3V spectrum. Main absorption lines are labelled. Shaded
vertical bands show wavebands with incorrect background subtraction. Spectra are given in
arbitrary units as we do not use an absolute calibration.}
\label{fig:1r22}
\end{figure}

According to our estimates, the spectral type of 1r and 11r is B1V up to 0.5
subtype; that of 4r and 5r is intermediate between B1V and B2V; and that of
6r and 22 is B5--B6V with an accuracy of 1 subtype. (We adopted the spectra
of MK standards from~\citealt{gray}.) All the obtained spectra
are shown in Fig.~\ref{fig:1r22}. According
to~\citet{rac74}, the spectral types of stars 1r, 5r, and 11r are
B1--2V, B3--5V and B2III, respectively. However, in spectra of B3--5V-type
stars conspicuous Mg{\sc II} 4481\,\AA\ absorption should be
present~\citep{gray}, while it is barely seen in the spectrum of star 5r
(Fig.~\ref{fig:1r22}) that was obtained with a substantially higher
dispersion than the spectrum obtained by~\citet{rac74}. \citet{rac74}
estimated the luminosity class of the star 11r as III based on smaller widths
of its Balmer absorption lines. However, the widths of these lines in our
spectra of stars 11r and 5r are nearly the same (Fig. \ref{fig:1r22}).
Furthermore, one should take into account the presence of Balmer emission
lines in the gas-dust nebulae surrounding the considered stars which can
distort the actual appearance of stellar absorption lines.

Approximating the total interstellar and circumstellar extinction by the
normal interstellar extinction law allows estimating the colour excess
$E(B-V)$ from the spectral energy distribution (SED). The colour excesses of
the investigated stars differ slightly from each other, being equal to 1.1
mag for 4r and 11r, 1.3 mag for 1r, 1.4 mag for 5r and 22, and 1.5 mag for
6r. The relative continuum flux difference between the
dereddened SEDs of all the studied stars and the SEDs of
standard stars of the corresponding spectral
types~\citep{val04,sil92} does not exceed 10 per cent at
all wavelengths in the considered spectral band.

Red spectra (up to $\approx 7300$ {\AA}) were taken only for stars 1r and 4r.
In the case of the star 1r the red spectrum complements rather well the
spectrum covering the 4400--5580 {\AA} wavelength interval and yields the
extinction parameters which do not differ substantially from those inferred
from the bluer spectrum. On the other hand, the continuum of the 4r star is
reproduced much better assuming a non-standard total-to-selective extinction
ratio of  $R_{\rm v} = 4.0$, which agrees with the $R_{\rm v}$ estimate
presented in the previous section. The colour excess in this case is $E(B-V)
= 1.05$ mag.

\section{Interstellar medium in the  vdB~130 area}

\subsection{Molecular cloud and ionized hydrogen in the vicinity of the cluster}

\begin{figure*}
\includegraphics[width=\linewidth]{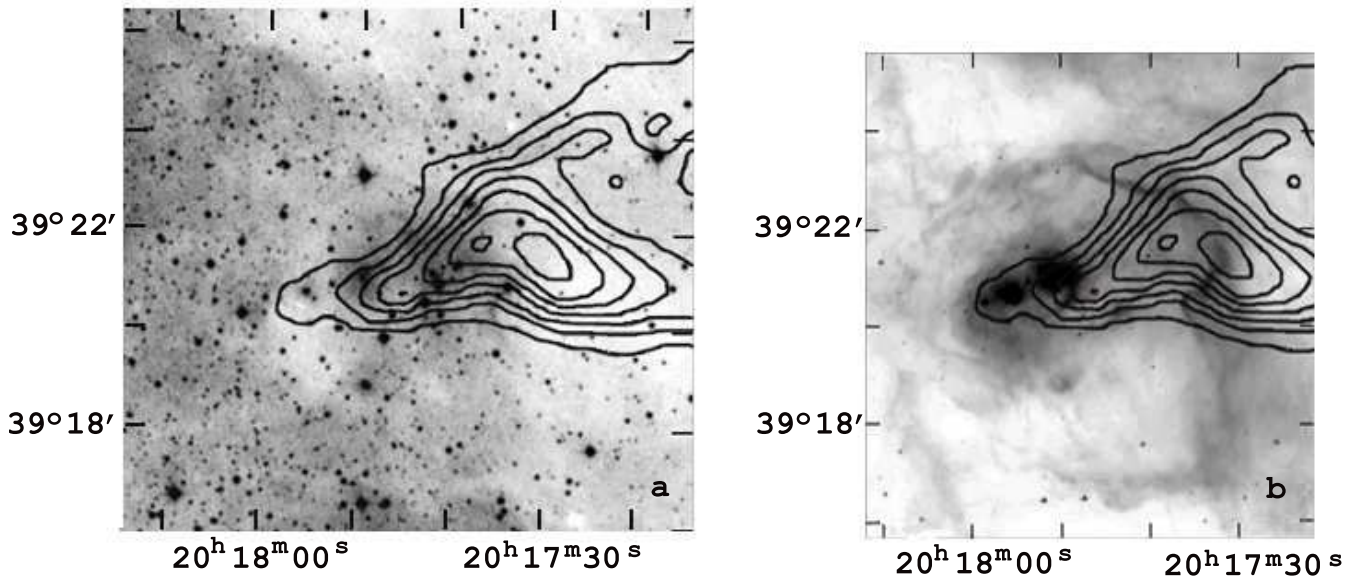}
\caption{VdB 130 region with contours of the `head' of the CO cloud
\citep{sch07} superimposed onto the E-DSS2 map (a) and onto the $8.0\,\mu$m
image (b).}\label{fig:cloud-head}
\end{figure*}

VdB~130 is located in projection near the north-western border of an extended
optical, radio and IR shell around Cyg OB1
(Fig.~\ref{fig:Cygpart})\citep{loz88, loz90, sak92}. This supershell is also
clearly discernible at all wavebands observed with the \textit{Spitzer}
space telescope \citep {hor09} and at large-scale CO distribution
\citep{sch07}. The distance to the association is 1.5 kpc \citep{gar92,
sit96}. From the analysis of the cluster stellar population, we estimate that
within errors Cyg OB1 and vdB~130 are located at the same distance (see
sect. 3.1).

The studied region is projected onto a cometary molecular cloud
(Fig.~\ref{fig:Cygpart}b, \ref{fig:cloud-head}a) having a length of $\simeq
0.4^{\circ}$ \citep[cloud A in][]{sch07}. The cloud is stretched along the
galactic latitude, with its `head' protruding inside the Cyg OB1 supershell.
The shape of the CO cloud hints that its `head' is actually a pillar
on the inner supershell border. Some stars of vdB~130 are seen in the region
of the cloud head, and all the other stars surround it
(Figs.~\ref{fig:vdB130}, \ref{fig:cloud-head}a). The CO emission from the
cloud spans the velocity interval $V_{\mathrm{LSR}}\sim-1\div5 \kms$
\citep{sch07}. The fact that the median line-of sight velocity for 34 out of
70 Cyg OB1 stars is $4 \kms$ with a standard deviation of $8.9 \kms$
\citep{sit03} suggests that the association and the cloud are close
neighbours in space. The presence of a reflection nebula and considerable
scatter of interstellar extinction estimates toward cluster stars can also be
an indirect indicator of a link between vdB~130 and the CO
cloud~\citep{rac74}.

The differential reddening could also be possible if the
cluster is in the background of the cloud, but the extinction
values found above are too low for this to be true. Indeed the
minimum $E(B-V)$ in the cluster area is 0.79 mag, which
corresponds to $A_V$ only of the order of 2.5 mag. On the other
hand, some stars, projecting onto the cloud head, have colour
excesses of the order of 1.3--1.5 mag. Assuming $R_V \sim 4 - 5$
(see Section~3), the extinction toward these stars generally
agrees with the estimates from \citet{sch07} for their Cloud~A
(5--8 mag).

Our FPI observations can be used to find kinematic evidence in favour of the
physical connection between the molecular cloud and ionized hydrogen in the
cluster outskirt. The data exhibit single-component H$\alpha$ profiles
toward vdB~130 with peak line-of-sight velocities ranging from 0 to
$15\kms$. These velocities characterize gas whose motions are mostly due to
Galactic rotation and streaming motions caused by spiral density waves.
To analyse characteristic systematic motions of ionized
hydrogen in the studied region, we use a `diagram of occurrence' of
line-of-sight velocities  (see, e.g.~\citealt{ark13} and references therein).
Characteristic line-of-sight velocities of the ionized gas
$V_{\mathrm{LSR}}$ in the region are defined as average values at half-maximum of the
corresponding diagram together with root-mean-square errors of their
estimates. The obtained interval of the most commonly occurring ionized gas
velocities toward vdB~130 is $\Delta V_{\mathrm{LSR}}\sim3 - 9 \kms$ (based
on 24 velocity measurements out of 26) and is shifted in the positive
direction relative to velocities of CO emission
$V_{\mathrm{LSR}}\sim-1\div2.5 \kms$ in the head of the cloud (see Fig.~2 in
\citealt{sch07}).

\begin{figure}
\includegraphics[width=\linewidth]{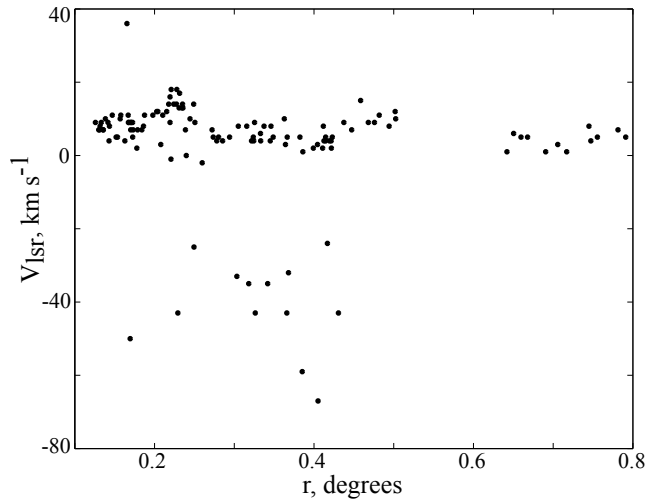}
\caption{Position-velocity diagram for the area shown by the white rectangle
in Fig. \ref{fig:Cygpart}. The star HD 193595 is located at the coordinate origin, and
the molecular cloud is located at the distance of $0.^{\circ}4$ from the star.}\label{fig:pv}
\end{figure}

Note that in the tail of the cometary molecular cloud (about $5 - 12$ arcmin
westward of vdB~130), where the influence of cluster stars should be less
pronounced, \HII\ velocities ($V_{\mathrm{LSR}}\sim3 - 6 \kms$) do indeed
coincide with CO velocities in the same area ($V_{\mathrm{LSR}}\sim 2 - 5
\kms$). This is an evidence in favour of the physical connection between
the CO cloud and ionized hydrogen in its environment. Minor
differences in $V_{\mathrm{LSR}}$ of molecular and ionized gas, observed in the
cloud head and in its vicinity, may indicate the influence of winds and radiation
from vdB~130 stars onto the surrounding gas.

Inside the supershell around Cyg OB1, profiles with multiple peaks are
observed. Specifically, in addition to the main component, they exhibit weak
high-velocity features in the wings of the H$\alpha$ line. These shifted
components reveal local high-velocity gas motions caused by stellar winds
and/or supernova explosions (with a possible contribution from distant \HII\
regions in the Perseus arm at large negative velocities).

\begin{figure*}
\includegraphics[width=0.48\linewidth]{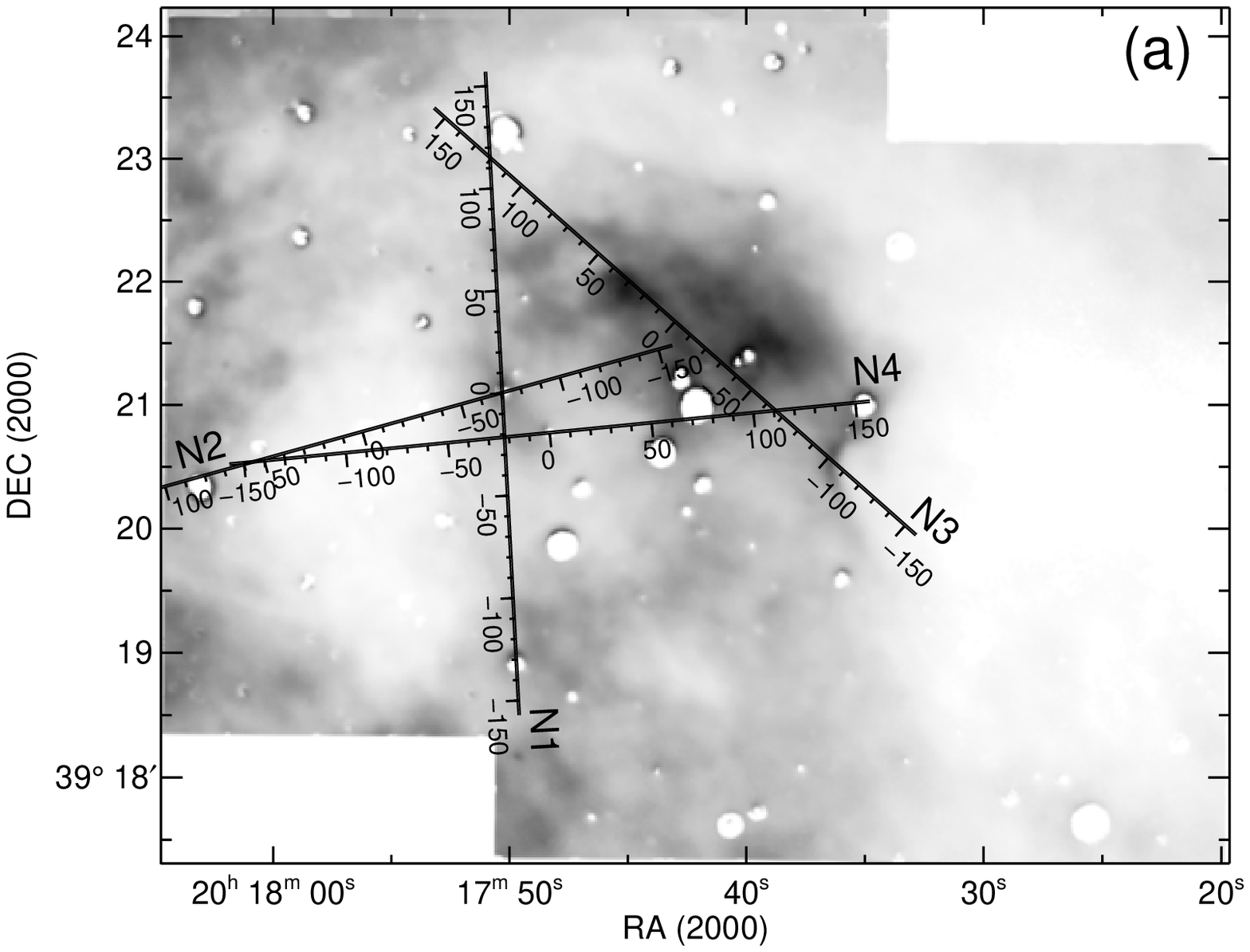}~
\includegraphics[width=0.48\linewidth]{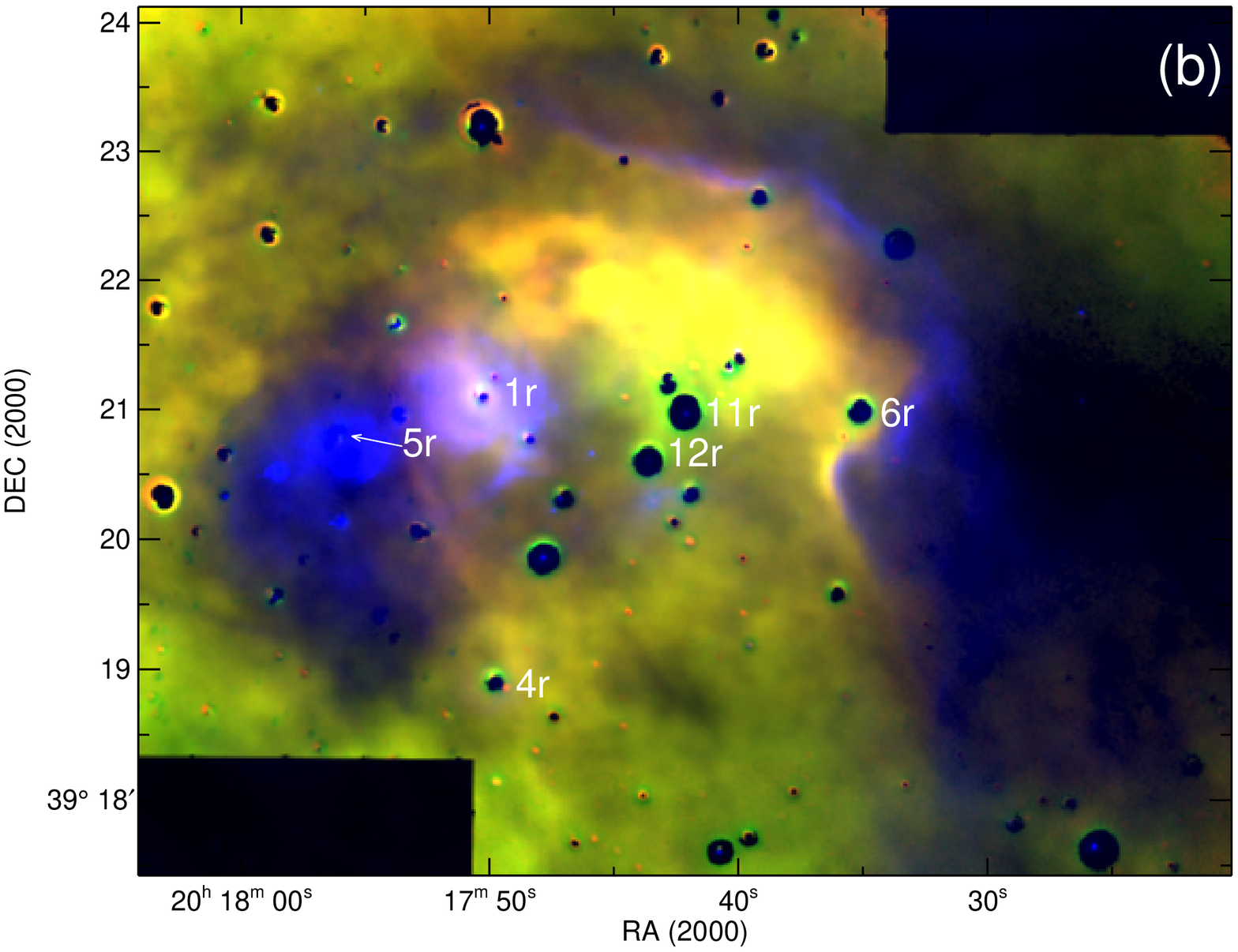}

\includegraphics[width=0.48\linewidth]{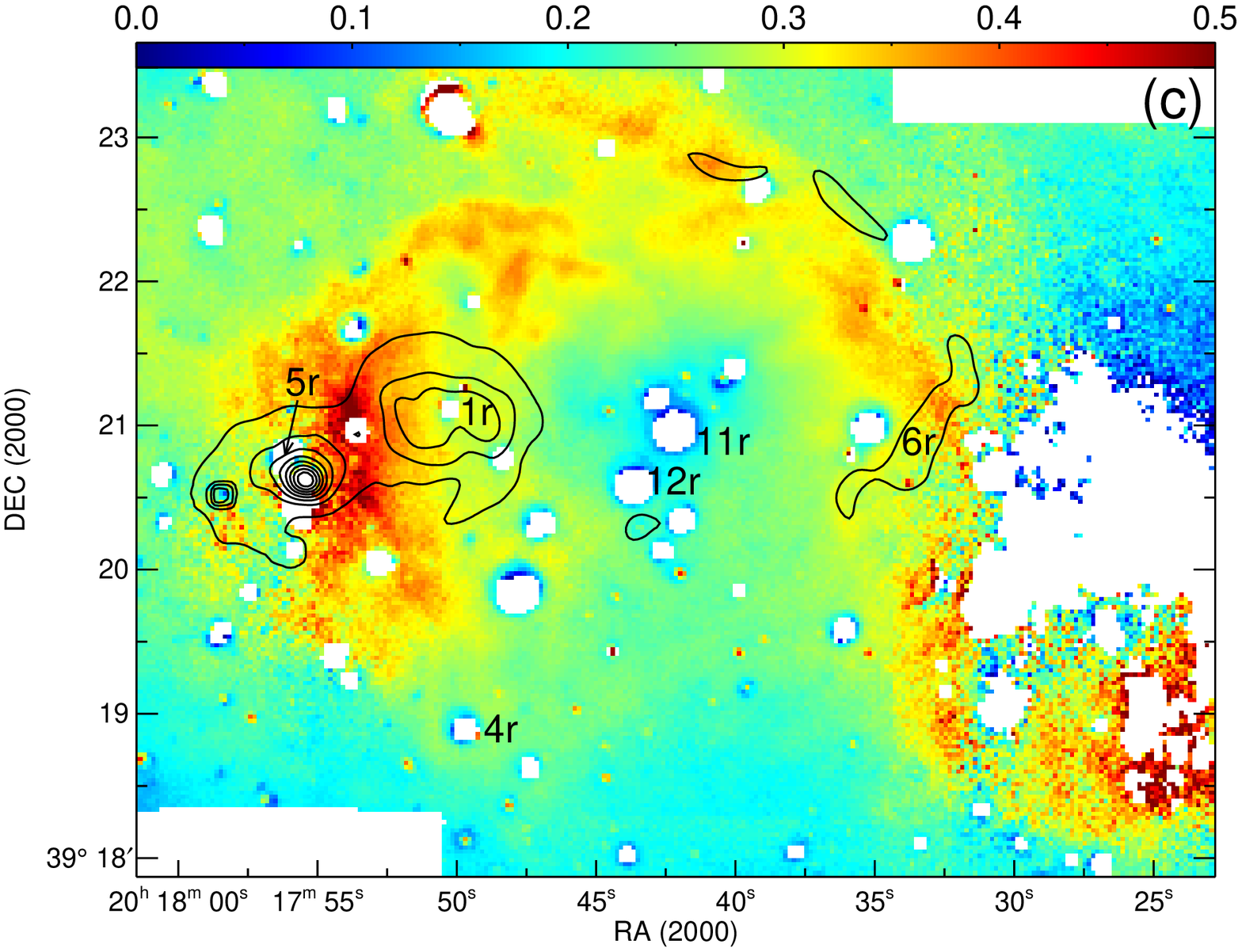}~
\includegraphics[width=0.48\linewidth]{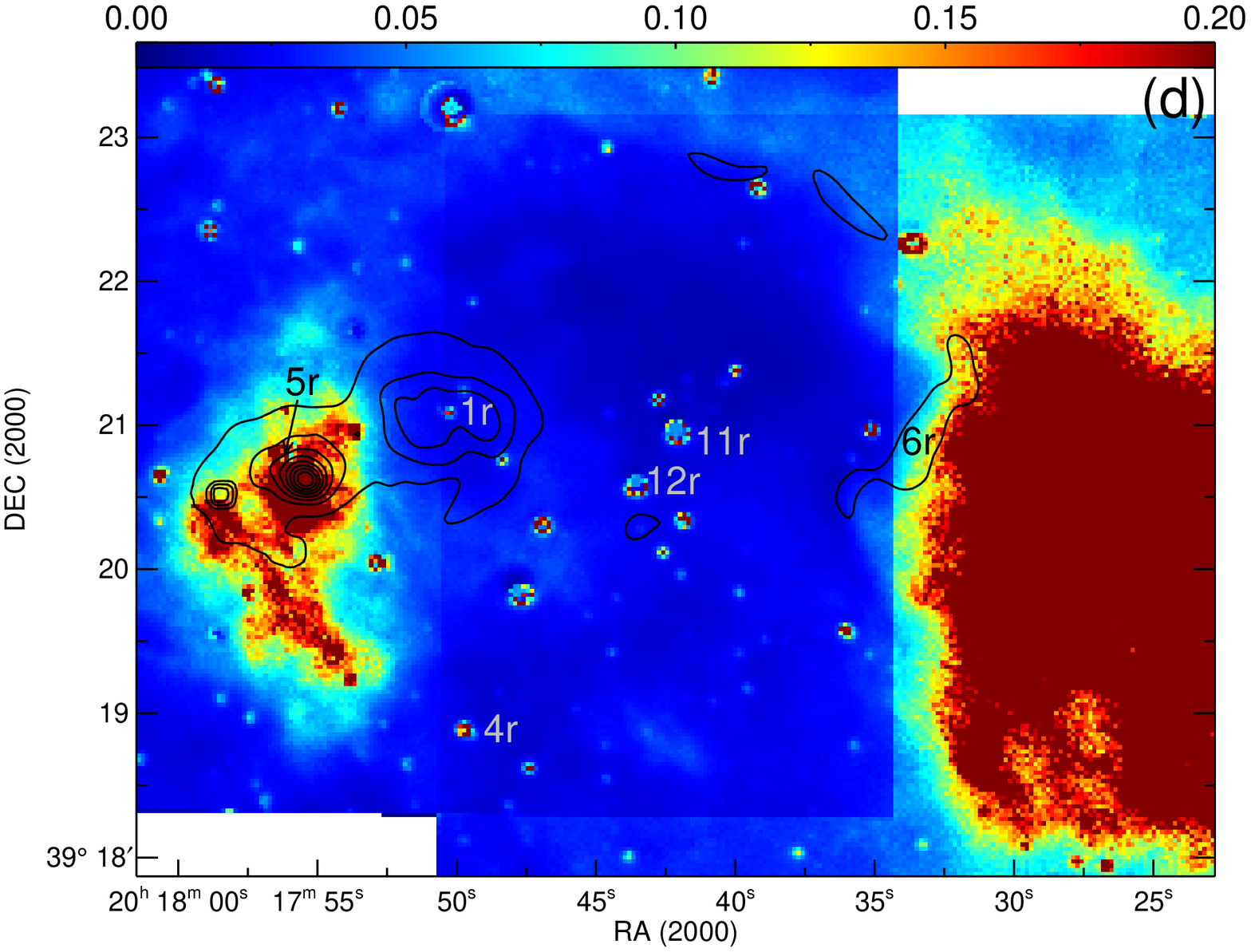}


\caption{Maps of ionized gas in the direction of vdB~130.  Bright stars are
masked in panels a) and c). Numbers indicate vdB~130 stars. (a) An \Ha\,
image and slit positions used for our spectral observations. (b) A
false-colour RGB-image where red colour corresponds to \SII\ emission, green
colour corresponds to \Ha\ emission, and blue colour indicates the flux
distribution in the {\em Spitzer} $8\,\mu$m band. (c) A map of
\SII6717,6731/\Ha\, flux ratio with overlaid isophotes in $8\,\mu$m band. (d)
A map of uncertainties  of the \SII6717,6731/\Ha\, flux ratio shown in the
panel (c) with overlaid isophotes in $8\,\mu$m band. }\label{fig:siitoha}
\end{figure*}

The molecular cloud can be affected by UV radiation and stellar
winds from at least two association members, namely, HIP100173 (O7e) and
HD193595 (O8), located approximately along the extension of the cometary
cloud axis, in 16~arcmin (7~pc) and 27~arcmin (12~pc) from the considered
region (Fig.~\ref{fig:Cygpart}). These stars are the nearest members of the
Cyg~OB1 association with strong stellar winds \citep{bla89,gar92}. However,
the entire vdB~130 star-gas-dust complex is located close to the boundary of
the supershell surrounding Cyg~OB1. Therefore, the projected expansion
velocity of the shell should be close to zero in the considered region.
Indeed, as we pointed out above, spectra taken in the vicinity of vdB~130
exhibit mostly single-peaked H$\alpha$ profiles.

In Fig.~\ref{fig:pv} we show the PV-diagram, i.e., the plot of \HII\
line-of-sight velocities vs. the projected distance to the HD193595 star. (The
area for which this PV-diagram is constructed is shown with a white
rectangle in Fig.~\ref{fig:Cygpart}a.) High-velocity \HII\ motions are
evident within $0.^{\circ}4$ from HD193595. This separation corresponds to
the distance from the star to the outer boundary of the supershell and to
vdB~130.

Thus, the cometary shape of the cloud, the close spatial location of the
cloud and the cluster, highly non-uniform reddening of vdB~130 stars, and the
distribution of line-of-sight velocities of ionized gas in the cloud
neighbourhood are all indicative of a possible physical connection between
the CO cloud, Cyg~OB1, and vdB~130.

However, the exact nature of this connection is not immediately clear.
CO emission implies relatively high density and correspondingly high extinction,
so that some cluster stars are at least partially embedded in the
absorbing material with $A_{\rm V}$ of the order of 3--6 mag.
However, the CO emission intensity in the direction of vdB~130 implies
higher extinction \citep{sch07} (see above) so that at least
some CO-emitting gas should be located behind the cluster. It is also
possible that some CO emission comes from the dense material related to
infrared features discussed below.

\subsection{Emission spectrum of the interstellar medium toward vdB~130}\label{sec:ism}

Figure~\ref{fig:siitoha} shows various \SII 6717,6731\ and \Ha\ emission maps
of the area surrounding vdB~130, based on our observations performed with
the 6-m telescope of the SAO RAS. An \Ha\ image is shown in
Fig.~\ref{fig:siitoha}a. In Fig.~\ref{fig:siitoha}b the \Ha\ map is combined
with a \SII\ map and an $8\,\mu$m \textit{Spitzer} map.

\begin{figure*}
\includegraphics[width=\linewidth]{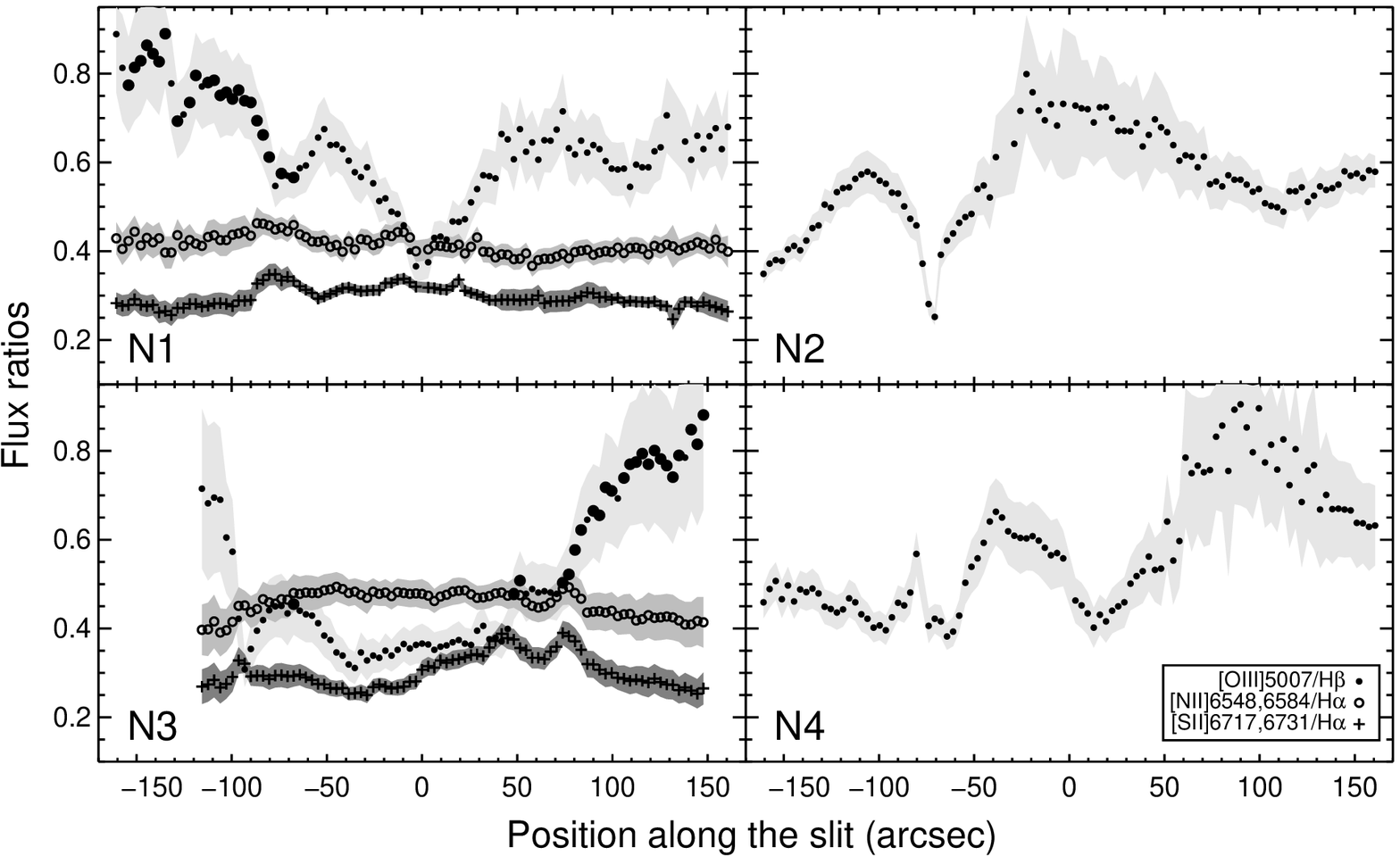}
\caption{Profiles of the \OIIIHb, \NIIHa, and \SIIHa\ flux ratios along four
slits shown in Fig.~\ref{fig:siitoha}a. The filled grey areas indicate the
uncertainties of the measured flux ratios. The \OIIIHb\ ratio is shown by the
lighter shade of gray and that of the \SIIHa\ ratio is shown by the darker
shade. The larger filled circles indicate the regions where we infer the
combined emission mechanisms (see text).}\label{fig:fluxes}
\end{figure*}

In the interstellar medium surrounding stars 10r, 11r, 12r, and 13r a
well-defined stratification of \Ha\ and \SII\ emission can be seen, which is
typical of \HII\ regions. Specifically, a `green region' dominated by \Ha\
emission surrounds the stars mentioned above in the centre and the
lower part of Fig.~\ref{fig:siitoha}b.
(Blue halos around stars Fig.~\ref{fig:siitoha}c are wings of a point spread function in H$\alpha$).
In the north-western part a `yellow region' is located, where both \Ha\ and
\SII\ lines are bright. (Note that a similar region in the south may be
partly obscured by the material responsible for the cluster reddening). The
periphery of the H$\alpha$ emitting area is dominated by `reddish' regions with enhanced
\SII 6717,6731/\Ha\ line intensity ratio. Dark dust clouds are located to the
west of the star 6r and to the east of the star 1r and around 5r, and also in the northern part of the region (these regions will be
discussed in Section 4.3).

The stratification of the emission is evident in the image showing the \SII
6717,6731/\Ha\ line intensity ratio (Fig.~\ref{fig:siitoha}c), which is traditionally used as a shock
indicator. At an average electron density $n_e \simeq 50-100$~cm$^{-3}$, as derived from the \SII$6717/6731$ flux ratio,
$I($\SII$6717,6731)/I($\Ha$)> 0.4$ is an evidence of a substantial
contribution from the emission of gas behind the shock front (see, e.g.,
\citealt{allen}). In Fig.~\ref{fig:siitoha}c regions with shock signatures
are shown in red. (Note that the \SII 6717,6731/\Ha\ flux ratios may be
underestimated because the image of the region actually combines \Ha\ and
\NII 6548,6584 lines (see Section 2), and this further supports our
conclusion about the possible presence of shocks.)

The most expressive shock manifestation is an X-shaped structure in the
eastern part of the region between stars 1r and 5r (Fig.~\ref{fig:siitoha}c).
However, high \SII/\Ha\ intensity ratio in this area may also be due to low
intensities of the lines resulting in large measurement errors. To separate
the two factors (shock excitation and observational errors), in
Fig.~\ref{fig:siitoha}d we show a map of uncertainties of the derived
\SII/\Ha\ intensity ratio at each pixel. The comparison of
Fig.~\ref{fig:siitoha}c and Fig.~\ref{fig:siitoha}d shows that the eastern
part of the X-shaped structure and the western part of the observed region
are characterized by significant uncertainties of the \SII/\Ha\ ratio and
that the enhanced \SII/\Ha\ ratio in these areas most likely arises due to
very low intensity of the lines. Hence only the western part of the X-shaped
structure can be considered as an indication of a shock front.

The spectroscopic observations of the area confirm that gas emission in the
vdB~130 region is actually a combination of photoionized gas emission and (in
some directions) shocked gas emission. The \NII6548,6583/\Ha\ line ratio
along the slits (Fig.~\ref{fig:fluxes}) exhibits high values in the same
regions where the \SII/\Ha\ line ratio is high. The average value of the
\NII6548,6583/\Ha\ line ratio is 0.40--0.45. The \OIII5007/H$\beta$ ratio
along the slits varies from 0.3 to 0.9. Local spikes of this
ratio in Fig.~\ref{fig:fluxes} may be associated with individual stars,
e.~g., 4r (slit N1, --70...--150) or 11r and 12r (slit
N4, 50--150) (see also Fig.~\ref{fig:siitoha}a). The enhanced
\OIII5007/H$\beta$ ratio in the region of the dust cloud to the east of the
star 1r (slit N2, -70...70) may be due to large measurement errors of weak
lines, as in the case of the \SII/\Ha\ ratio.

\begin{figure}
\includegraphics[width=\linewidth]{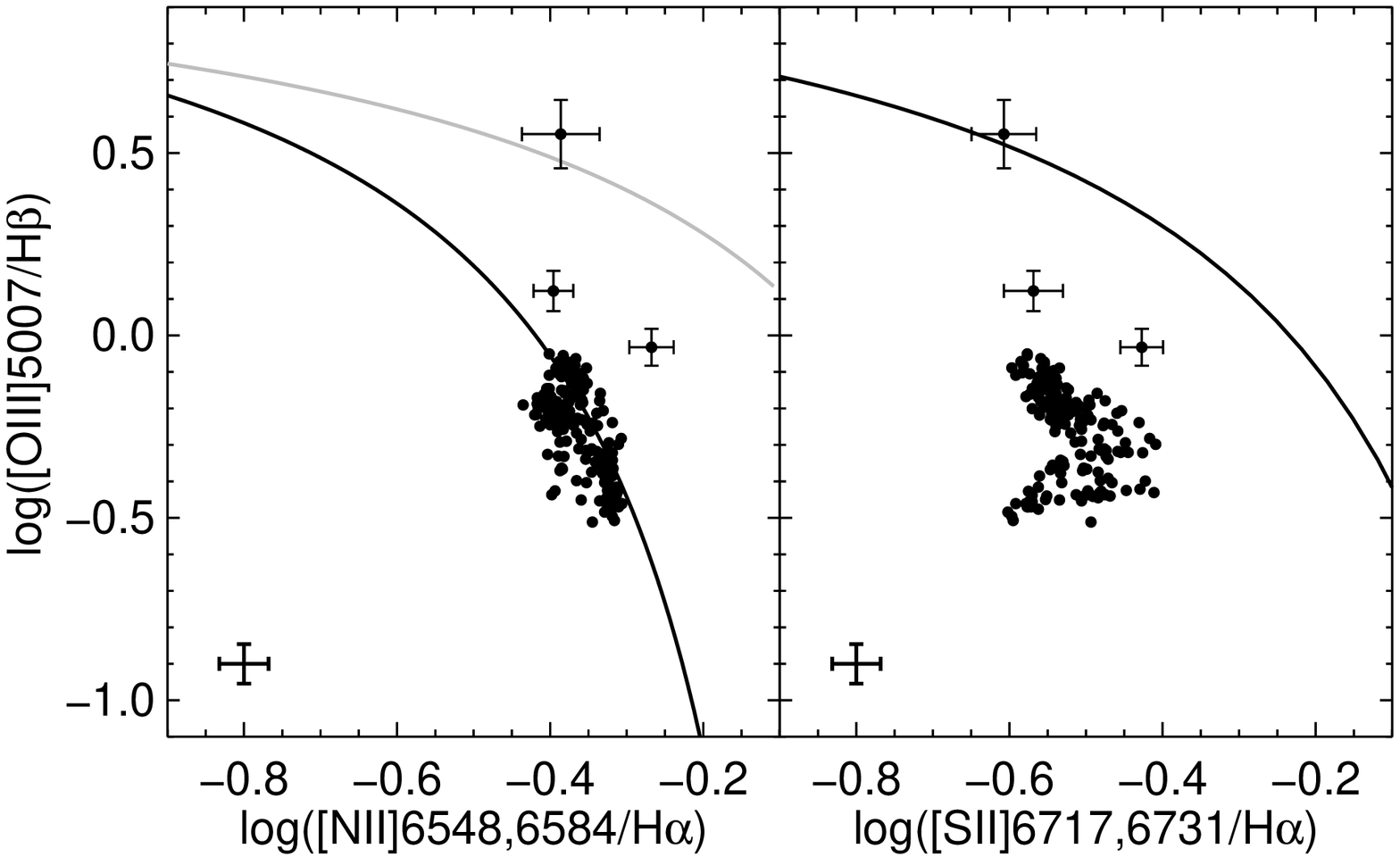}
\caption{Diagnostic diagrams of the relations $\log\mathrm{([OIII]/H\beta)}$
to $\log\mathrm{([NII]/H\alpha)}$ (left) and to $\log\mathrm{([SII]/H\alpha)}$
(right), observed with the spectrograms N1 and N3 (Fig.~\ref{fig:siitoha}a).
Solid lines, according to the models of \citet{Kauf03} and \citet{Kewley2001} and their
parametrization by \citet{Kewley2006}, separate regions of pure
photoionization excitation (below the black line), shock excitation (above
both lines), and combined contribution of both mechanisms (between grey and
black lines in the left panel). Typical measurement errors are shown in the
bottom left corner of each panel.}\label{fig:bpt}
\end{figure}

In Fig.~\ref{fig:bpt} we relate our observed diagnostic diagrams
$\log\mathrm{([OIII]/H\beta)}$ vs. $\log\mathrm{([NII]/H\alpha)}$ and
$\log\mathrm{([OIII]/H\beta)}$ vs. $\log\mathrm{([SII]/H\alpha)}$ to
theoretical photoionization models of \cite{Kewley2001} and \cite{Kauf03}. Black lines
mark the upper boundaries for possible line intensity ratios in the case of
purely photoionization excitation mechanism. A grey line in the left panel
separates a domain with \NII/\Ha\ and \OIII/\Hb\ ratios typical of shocks
(above the grey line) from the domain with the combined contribution of both
mechanisms (between the black and grey lines). Evidently, in the studied
region we observe mostly photoionization line excitation. However, a number
of data points in the left panel fall inside the `composite' domain. In
Fig.~\ref{fig:fluxes} these points are marked by bigger circles on the
profiles of the  \OIII/\Hb\ ratio. Comparing Fig.~\ref{fig:fluxes} (slits N1
and N3) and Fig.~\ref{fig:siitoha}a, we see that domains with combined
emission mechanism coincide with the shell that is traced by the enhanced
\SII/\Ha\ ratio (Figs.~\ref{fig:siitoha}c,d), indicative of a possible
contribution from shocks. In particular, an analysis of spectroscopic
observations confirms the combined emission mechanism in the western segment
of the X-shaped structure, which is a part of the shell-like pattern visible
in the \SII/\Ha\ distribution (see slit N1, -80...-70 in
Fig.~\ref{fig:fluxes}).

Thus, we conclude that an \HII\ region with possible shock signatures at its
periphery is observed around a group of vdB~130 members.
Unfortunately, we were able to perform a comprehensive analysis only for the
N1 and N3 slits. The wavelength ranges for the other two slits
include only \OIII\ and \Hb\ lines.

We also used spectra N1 and N3 to estimate the gas metallicity. The \OIII
4363 line that is needed for estimating $T_e$ was too weak and we therefore
could not apply the so-called direct method, which requires the knowledge of
the electron temperature. An analysis of the gas metallicity based on the
empirical method from \citet{abund} calibrated by bright \NII\ and \SII\
lines yielded a close-to-solar oxygen abundance of $12+\log\mathrm{(O/H)} =
8.6 \pm 0.15$.

\subsection{Dust emission in the vdB~130 area}\label{sec:dust}

\begin{figure*}
\includegraphics[width=0.45\linewidth]{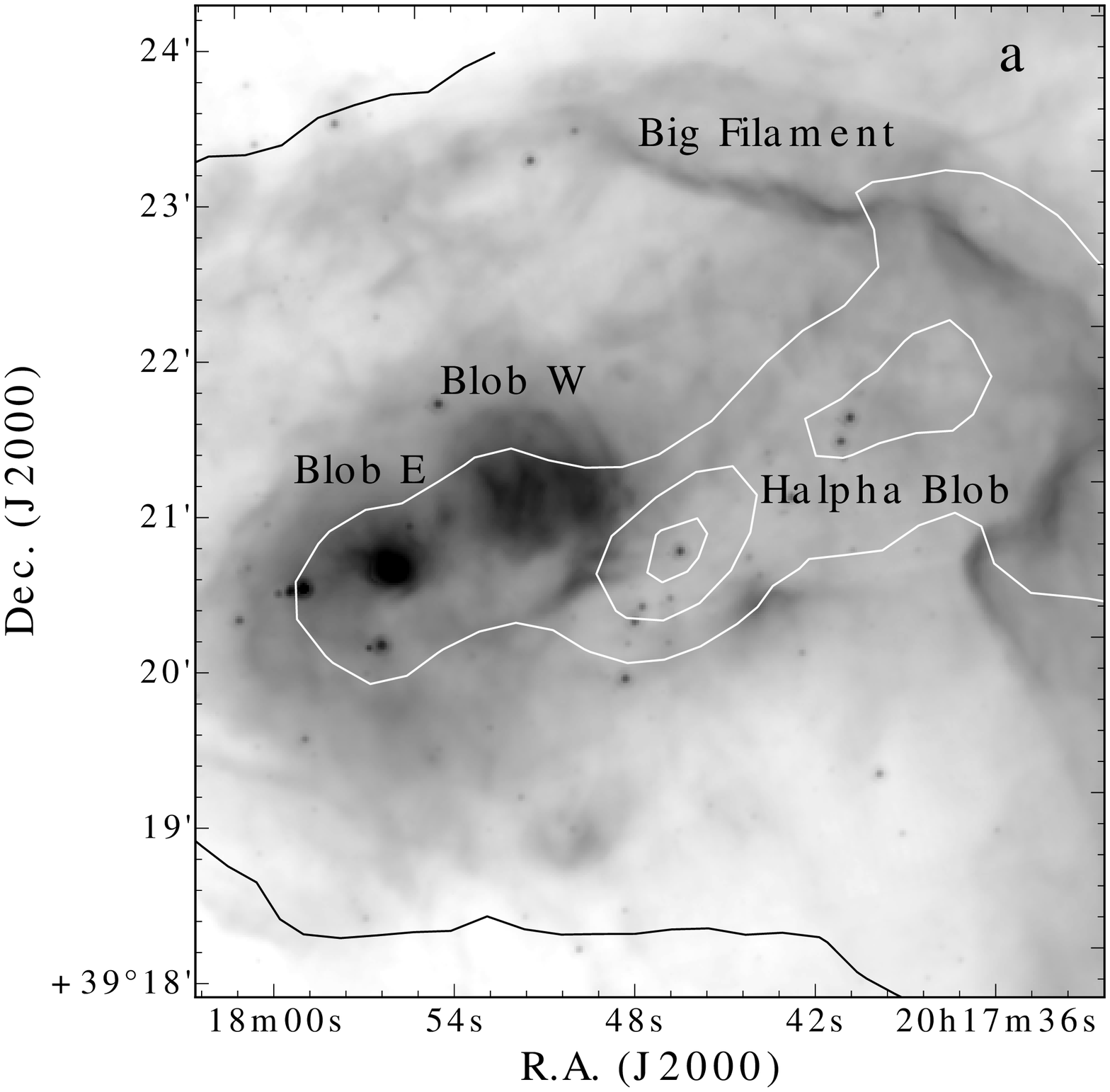}
\includegraphics[width=0.45\linewidth]{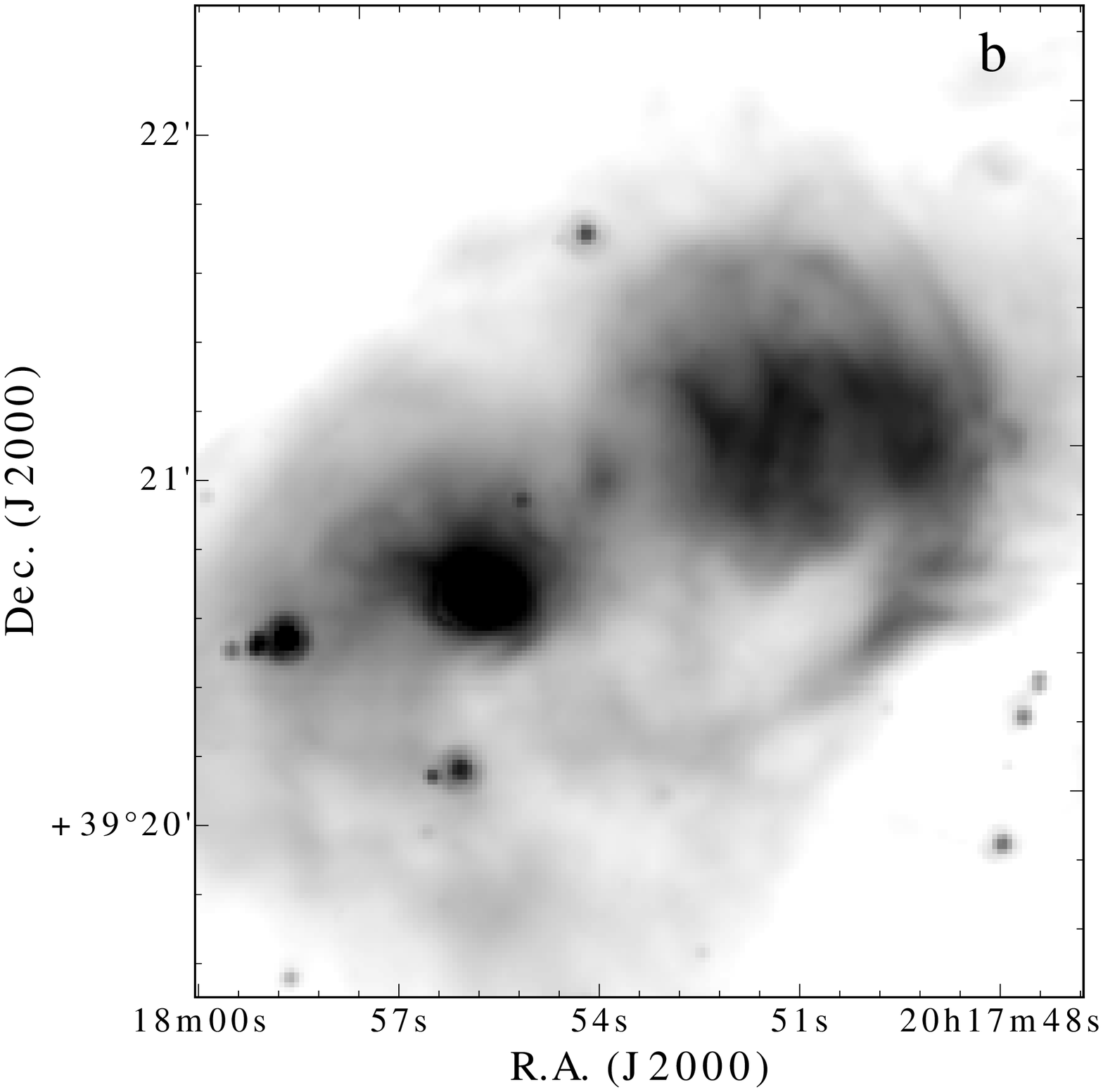}

\includegraphics[width=0.45\linewidth]{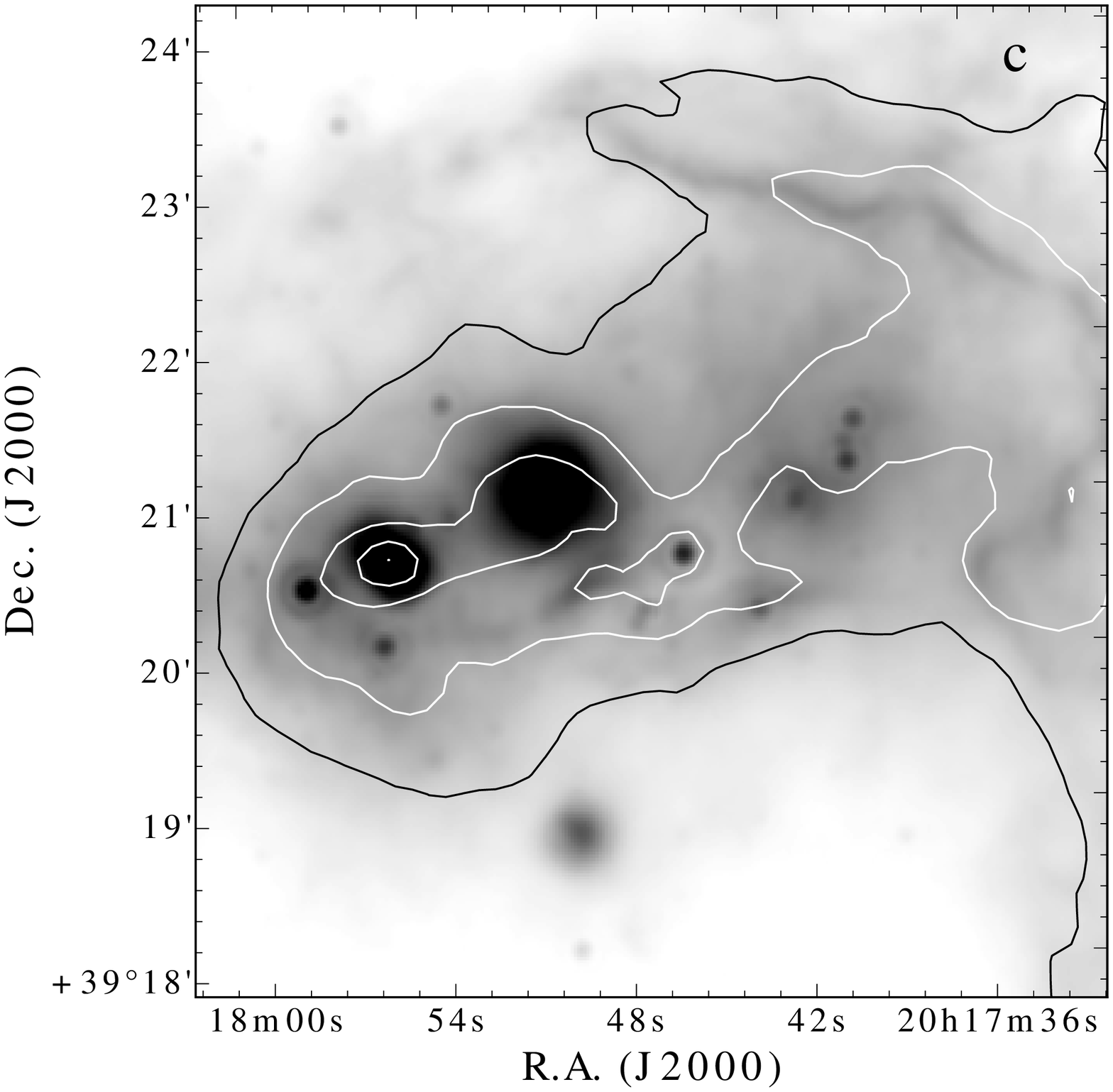}
\includegraphics[width=0.45\linewidth]{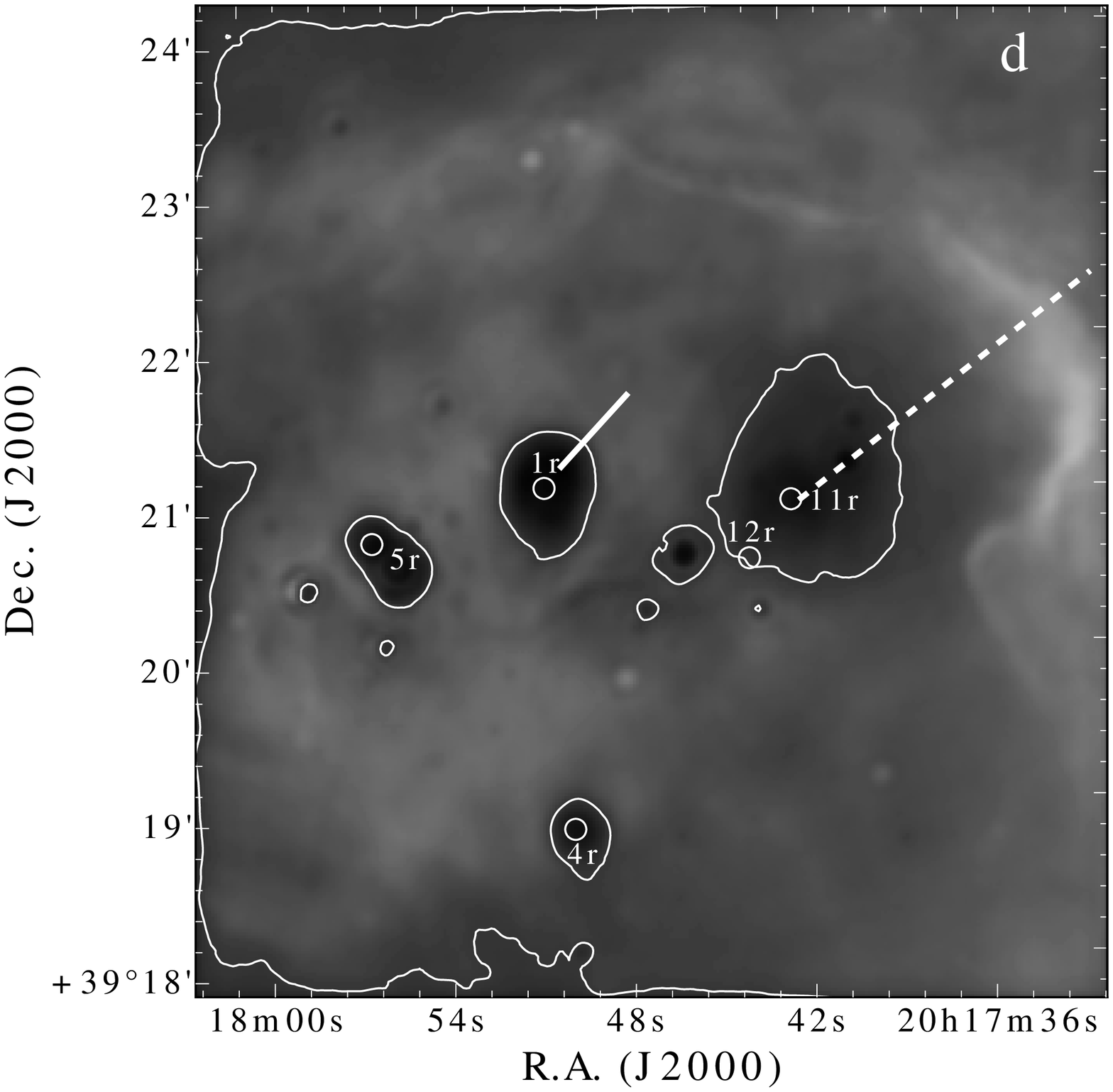}

\caption{(a) A map of the studied region at 8\,$\mu$m with overlaid contours
of 500\,$\mu$m emission. (b) An expanded area of the 8\,$\mu$m map around the
infrared blobs. (c) A map of the studied region at 24\,$\mu$m with overlaid
contours of 160\,$\mu$m emission. (d) A map of the ratio of fluxes at
8\,$\mu$m and 24\,$\mu$m. Dashed and solid lines indicate cuts for which flux
ratio profiles are presented in Fig.~\ref{lineratios}a and
Fig.~\ref{lineratios}b, respectively.} \label{IR}
\end{figure*}

Near-infrared images of the studied region in various {\it Spitzer} and {\it
Herschel} bands are presented in Figs. ~\ref{fig:Cygpart}b,
\ref{fig:cloud-head}b and \ref{IR}a--c. Infrared emission at 8\,$\mu$m and
24\,$\mu$m traces an unclosed shell with a diameter of $5 - 7$~arcmin (about
3 pc) and a thickness from 0.5 to 1~arcmin (about 0.4 pc), which encircles
most of the stars of vdB~130. The shell is open in the direction of Cyg~OB1
association toward the east and partially overlaps with the molecular cloud
toward the west (Fig.~\ref{fig:cloud-head}b). The brightest part of the shell
is further referred to as Big Filament. Inside the shell there is an extended
area of diffuse infrared emission with two blobs (hereinafter Blob~E and
Blob~W), easily discernible on both IRAC and MIPS (24\,$\mu$m) maps
(Fig.~\ref{IR}a,b,c). These blobs and the surrounding halo of bright infrared
emission coincide in projection with the head of the molecular cloud (Fig.
\ref{fig:cloud-head}b) and are embedded into the region emitting at
160\,$\mu$m, shown with contours in Fig.~\ref{IR}c. Blobs W and E have sizes
of $\sim1$~arcmin (0.4~pc) and $\sim0.5$~arcmin (0.2 pc), respectively.

In Fig.~\ref{IR}b we show an expanded area of the 8\,$\mu$m emission
map around the infrared blobs. Blob~W (western) is more extended and
represents a nearly round patch of IR emission, with a sickle-like filament
extending from its southern side. The fine structure with other filaments
within the blob is visible at 8\,$\mu$m (as shown in more detail in
Fig.~\ref{IR}b). The eastern blob (Blob~E) is more compact and actually looks
almost like a point source. These blobs and a sickle-like filament are also
visible on a 24\,$\mu$m map (Fig.~\ref{IR}c).

Appearances of 8 and 24\,$\mu$m images are similar but not identical which is
evident on the plot representing their ratio (Fig.~\ref{IR}d). To produce
this plot, we have convolved the 8\,$\mu$m map to the resolution of the
24\,$\mu$m map, using an appropriate kernel provided by \cite{ani11}. While
over most part of the studied region the $F_8/F_{24}$ ratio exceeds~1, there
are few patches where it is smaller than 1, down
to about 0.2. In Fig.~\ref{IR}d these patches are encircled by white contours
corresponding to $F_8/F_{24}=1$. Smallest values of $F_8/F_{24}$ are observed
toward Blob~E and Blob~W, but there is another quite extended region where
$F_8/F_{24}$ is less than~1, located westward of Blob~W, in the centre of a
structure delineated by a curved part of Big Filament.

A dissimilarity between 8\,$\mu$m and 24\,$\mu$m emission is a well-known
feature of the ISM both in our Galaxy \citep{bubble,bubblier} and in other
galaxies \citep{bendo}. In our Galaxy interiors of \HII\ regions tend to
be brighter at 24\,$\mu$m while their borders are observed as ring-like
structures on 8\,$\mu$m maps. It is tempting to assume that dark spots in
Figure~\ref{IR}d, where 8\,$\mu$m emission is suppressed relative to
24\,$\mu$m emission, indicate the presence of \HII\ regions. Below we
discuss whether or not other evidences may be presented in favour of this
hypothesis.

We start from the region encircled by Big Filament, which coincides with the
region of ionized gas emission discussed in the previous subsection. This
region is relatively faint at 8\,$\mu$m and is filled with diffuse 24\,$\mu$m
and H$\alpha$ emission (hereinafter we refer to it as H$\alpha$ Blob as
indicated in  Fig.~\ref{IR}a). Big Filament itself is bright at 8\,$\mu$m,
less bright at 24\,$\mu$m, and dark in H$\alpha$ emission. It is also clearly
visible on 70 and 160\,$\mu$m images, obtained with {\em Herschel}, and stays
discernible up to 500\,$\mu$m. The ring of infrared emission has been also
mentioned in \cite{sch07}.

\begin{figure}
\includegraphics[width=\linewidth]{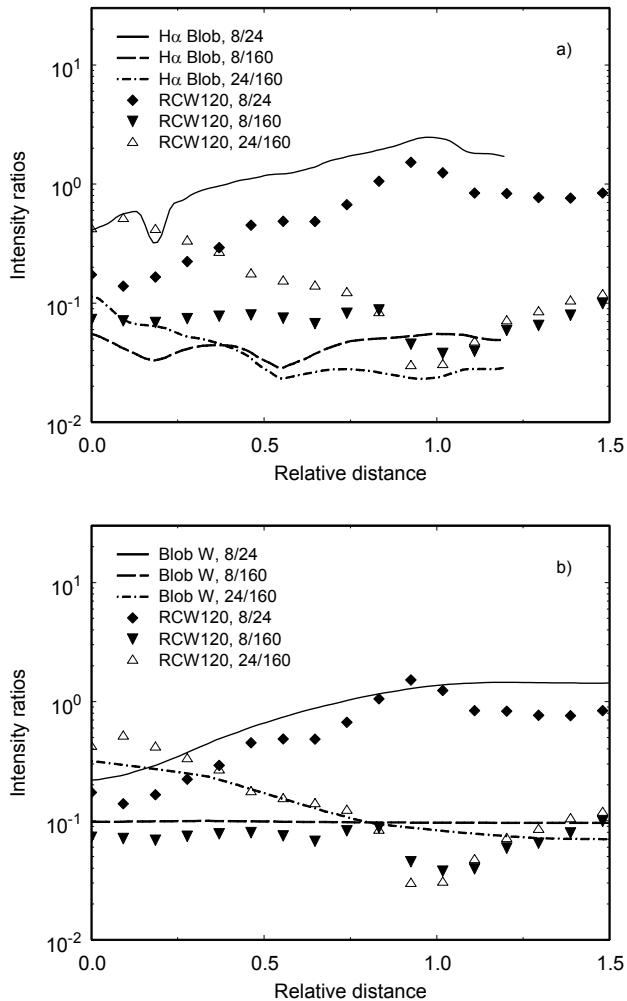}
\caption{Profiles of intensity ratio for H$\alpha$ Blob (a) and Blob~W (b).
For comparison analogous profiles are also shown for a representative cut
through the well-studied \HII\ region RCW~120.} \label{lineratios}
\end{figure}

The entire infrared emission pattern looks similar to Galactic \HII\ regions.
To emphasize this similarity, in Figure~\ref{lineratios}a we show how various
infrared flux ratios change with the relative distance from the region centre
along the cut shown in Figure~\ref{IR}d with a dashed line. Also shown in
Figure~\ref{lineratios}a are the same ratios in the \HII\ region RCW~120, one
of the best studied `infrared bubbles' in our Galaxy \citep{deh,and,kirsanova}.
For all the three ratios trends are similar. The $F_8/F_{24}$ ratio steadily
grows from the region centre (marked by the smallest value of $F_8/F_{24}$)
toward its border, while the $F_{24}/F_{160}$ ratio decreases along the same
cut down to 0.5 and then stays nearly constant. The ratio of $F_8$ flux to
$F_{160}$ flux exhibits the smallest variation staying approximately constant
over the entire region. Both 8\,$\mu$m and 24\,$\mu$m maps have been
convolved to a 160\,$\mu$m resolution using kernels from \cite{ani11}.

Blob~W is even more similar to RCW~120 in this respect as shown in
Figure~\ref{lineratios}b, with its interior being relatively
brighter at 24\,$\mu$m than H$\alpha$ Blob. Intensity ratios are
shown for the cut indicated with a solid line in Fig.~\ref{IR}d.
Blob~E is not resolved well enough for a detailed comparison, but
it is also faint at 8\,$\mu$m, bright at 24\,$\mu$m, and shows
little variation in the $F_8/F_{160}$ ratio. To ensure
that this similarity is not bound exclusively to RCW~120, we have
compared intensity ratios in our blobs with the corresponding
ratios in a few other infrared bubbles from the \cite{Churchwell}
catalogue and found the same distribution on the infrared fluxes.
Thus, flux ratios in {\em Spitzer\/} and {\em Herschel} bands in
all the three blobs shows spatial variations that are typical of
the \HII\ regions in the Galaxy.

\begin{figure*}
\includegraphics[width=0.4\textwidth]{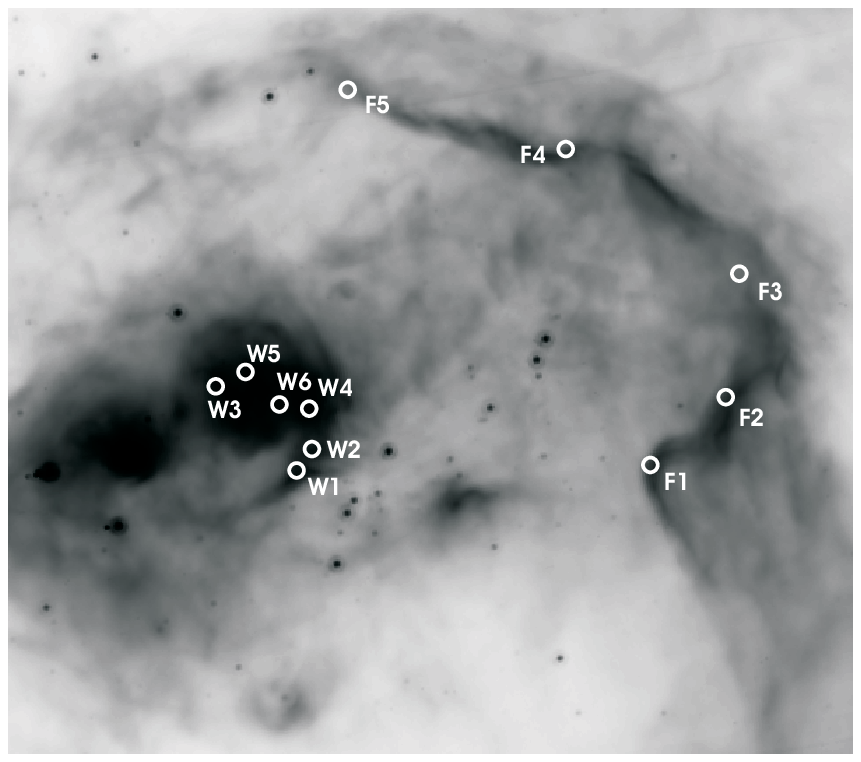}
\includegraphics[width=0.4\textwidth]{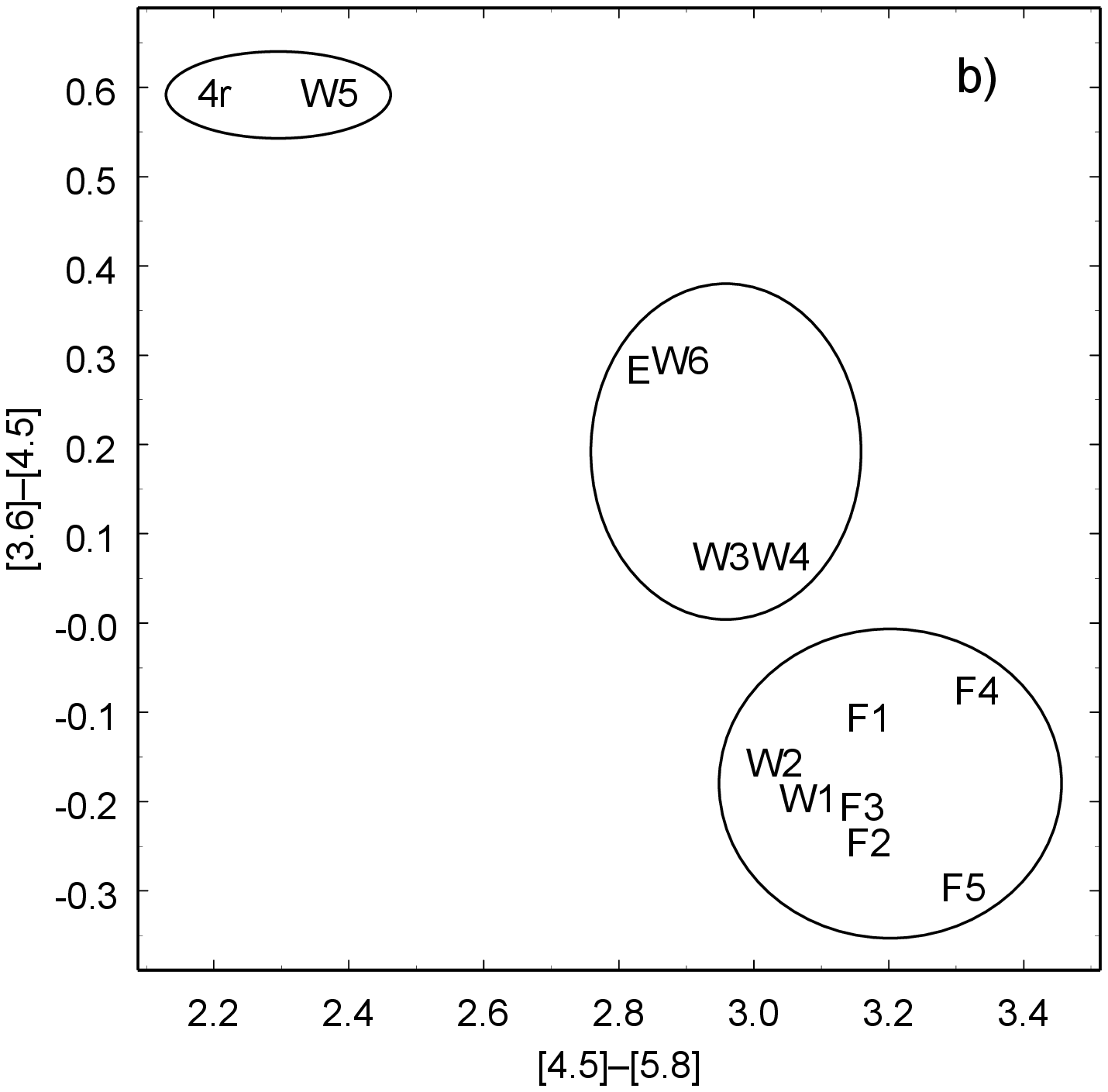}
\caption{(Left) Locations of the regions, for which the colour-colour diagram
is presented, on the $8\,\mu$m map. (Right) Colour-colour diagram for the
regions shown on the left panel. The regions {\it w3} and {\it w4} and the
regions E and {\it w6} have been moved apart slightly to avoid overlapping.}
\label{colours_reg}
\end{figure*}

We have also performed an aperture photometry in {\em Spitzer\/} IRAC
bands for several locations along Big Filament, along the sickle-like
feature in the southern part of Blob~W, and also for
Blob~E and a nebulosity surrounding the star 4r (with the central source
contribution subtracted). In each band surface brightness was measured
for a circular region with a diameter of 5 arcsec. Background emission was estimated
in extended regions with the lowest surface brightness within 100 arcsec from
the studied objects. The regions listed above are indicated in Fig.~\ref{colours_reg}a,
while their infrared colours [3.6]--[4.5] and
[4.5]--[5.8] are shown in Fig.~\ref{colours_reg}b.

All the regions can be separated into three groups according to their relation
to the domains described in \cite{ybarra2014}. The first group
comprises all the locations in Big Filament along with the {\it w1} and {\it
w2} regions in the sickle-like feature. These regions have negative
[3.6]--[4.5] colours and [4.5]--[5.8] colours exceeding~3. According to the
results of \cite{ybarra2014}, these colours are typical of a photon dominated
region (PDR) at a boundary of an \HII\ region with a moderate absorption
($A_V\sim(4-5)^{m}$) and $\log(G/n_{\rm H})$ of the order of --1. Here $G$ is
the FUV radiation field in units of the Draine field \citep{Draine1978}, and
$n_{\rm H}$ is the gas density. Our estimate for the electron density of the order
of 100 cm$^{-3}$ implies $G>10$. Note that in our picture all these regions are related to the boundary of the
\HII\ region around  vdB~130.

The second group consists of Blob~E and the {\it w3}, {\it w4}, and {\it w6}
regions, that belong to Blob~W. Their colours do not fall into the domain
considered in the work of \cite{ybarra2014}, with the [3.6]--[4.5] colour
being positive and the [4.5]--[5.8] colour being slightly below 3.
Extrapolating results of \cite{ybarra2014}, we may assume that absorption in
these directions is higher than $5^{m}$ and $\log(G/n_{\rm H})$ is lower than
--2. We cannot expect that $G$ is low in the region of bright near-IR
emission, so lower $\log(G/n_{\rm H})$ is probably caused by higher density
(consistent with higher absorption). Note that extinction values
inferred from \cite{ybarra2014} diagram agree with our estimates.

Finally, the third group in our sample includes the {\it w5} region and the
nebulosity around the star 4r. Location of the {\it w5} region nearly
coincides with the location of the star 1r. According to \cite{ybarra2009},
their colours ([3.6]--[4.5] $\sim0.6$ and [4.5]--[5.8] $\sim2.2-2.4$) can be
explained by the emission of the low-temperature shocked H$_2$ gas.

Thus, we may tentatively conclude that H$\alpha$ Blob, Blob~W, and Blob~E and
may be other regions with suppressed 8\,$\mu$m emission, also seen in
Fig.~\ref{IR}d (like a region around the star 4r), are actually \HII\ regions
surrounding young massive stars. In particular, the infrared emission
properties of H$\alpha$ Blob are entirely consistent with our conclusion on
the \HII\ region that surrounds the cluster vdB~130, based on the optical
spectra. Blob~W and Blob~E can be excited by the
stars 1r (B1V) and 5r (B1V-2V), respectively (Fig.~\ref{fig:vdB130}), while
H$\alpha$ Blob can be excited, among others, by stars 11r (B1V) and 12r,
which are members of the cluster vdB~130.

Being more extended and fainter in all infrared bands up to 160\,$\mu$m,
H$\alpha$ Blob is probably older than the other \HII\ regions in this area.
We note that Big Filament which borders H$\alpha$ Blob from the west,
partially encircles a ring-like feature, seen on the map of the
\SII/H$\alpha$ ratio (Fig.~\ref{fig:siitoha}b), which is also visible as an
absorption feature on the DSS image. The \SII/H$\alpha$ ratio is somewhat
enhanced along this ring-like feature relative to its interior, indicative of
a shock (Fig.~\ref{fig:siitoha}c). As we mentioned in the previous
subsection, in the eastern part of the region we observe a clear signature of
a shock. The sickle-like feature bordering Blob~W from the west may have been
formed by H$\alpha$ Blob expansion, when the ionization front has started to
engulf a parent clump for Blob~W.

The presence of \HII\ regions in the vicinity of vdB~130 stresses the
problem of its age overestimation. If H$\alpha$ Blob is indeed excited by the cluster star(s), their
interaction should have started quite recently, after the cluster itself had formed.
It must be noted that the stars 1r and 5r, which we propose as exciting stars for
Blob~W and Blob~E (as there are no other B stars in the blobs' vicinity),
may not be cluster members (see above). It is interesting that the three blobs are located
at the tip of a long chain of millimetre sources that extends to the west far beyond
Big Filament \citep{Motte2007}. This also hints that the CO cloud, also
visible in the dust continuum, may be a remain of a bigger structure that has been
destroyed by the expanding supershell. It is also possible that the interaction between the
pre-existing cloud and the expanding supershell has moved some molecular material in the
vicinity of the cluster B-stars, initiating the formation of H$\alpha$ Blob.

We may ask then whether Blob~W and Blob~E formation has been triggered by an
expansion of H$\alpha$ Blob. Their location hints at
some causal link with the shell, but the available information is definitely
not sufficient for any firm conclusion.

\section{Conclusions}

Detailed analysis of stellar populations and the interstellar medium in the
vicinity of the cluster vdB~130 was performed, using optical observations
taken with the 6-m telescope of SAO RAS and the 125-cm telescope of the
Crimean Laboratory of SAI MSU as well as archival data from \textit{Spitzer},
\textit{Herschel} Space Telescopes, 2MASS and UCAC4 projects.

The following results are obtained.

\begin{enumerate}
\renewcommand{\theenumi}{\arabic{enumi}.}
\item New revision of the stellar population of  vdB~130 revealed
36 stars with proper motions differing by less than $4 \myr$ from
the centroid determined by \citet{khar13} ($\langle\mu_{\alpha},
\mu_{\delta}\rangle \approx (-2.22, -4.39) \myr$). The sample
shows large scatter of colours (0.6 mag for faint stars) that can
be considered as an evidence for substantial differential
reddening across the cluster. The minimum colour excess $E(B-V)
\approx 0.79 \pm 0.02$~mag, and the apparent distance module
$(V-M_V)_{app} \approx 15.0 \pm 0.3$~mag, whereas in NIR $E(J-H)
\approx 0.27 \pm 0.02$~mag and $(K-M_{Ks})_{app} \approx 11.57 \pm
0.3$~mag. Since NIR data are corrupted by large and inhomogeneous
extinction in much lesser extent then in optics, we consider
cluster parameters obtained from 2MASS photometric data as more
reliable. Photometric data are consistent with young cluster age
ranging from 5 to 10 Myrs and distance of about $1.8 \pm 0.3$~kpc.

\item The physical relation between vdB~130, which is located near
the outer boundary of a supershell surrounding the Cyg~OB1
association, and the cometary molecular cloud with the size of
$0.^{\circ}4$ \citep{sch07} is corroborated by (1) similar
distances to the cluster and the association (1.5 -- 1.8~kpc),
which agree within the errors, and (2) the agreement of
line-of-sight velocities of the cloud
($V_{\mathrm{LSR}}\sim-1\div5 \kms$), Cyg OB1 stars, and hydrogen
ionized by these stars ($V_{\mathrm{LSR}}\sim3 - 9 \kms$). The
molecular cloud is located in the region influenced by the stellar
wind and radiation of Cyg~OB1 and vdB~130 stars.

\item Most vdB~130 stars are surrounded
by an IR shell having a diameter of 3 pc, which is visible at all
\emph{Spitzer} bands (Figs.~\ref{fig:cloud-head}b and \ref{IR}). The inner
part of the shell is partially filled with diffuse emission at 8 and
24\,$\mu$m, with 24\,$\mu$m emission being brighter than 8\,$\mu$m emission.
Such a relation between 8\,$\mu$m and 24\,$\mu$m is typical for Galactic
\HII\ regions, so we suggest that this shell marks the location of an
ionization  front in an \HII\ region, H$\alpha$ Blob, which is also visible
on our H$\alpha$ and \SII\ maps and is probably produced by the hottest stars
of vdB~130. At the eastern border of the \HII\ region encircled by the IR
filament signatures of a weak shock are identified.

\item There are at least two other \HII\ regions in the area,
Blob~W and Blob~E, which are more compact and mostly visible in infrared
maps. Blob~W and Blob~E are excited by B1V and B1V-B2V stars, respectively.
On the maps of the \SII/H$\alpha$ ratio Blob~W is delimited from the east by
an arc-like feature which may be produced by the
expansion of H$\alpha$ Blob.

\item Blob W and Blob E may represent the latest step in a star formation
sequence in the area related to the pillar-like CO cloud, but in general the star
formation history across the area seems to be quite complicated.

\end{enumerate}

\section*{ACKNOWLEDGEMENTS}

We are grateful to the anonymous referee for her/his extremely thorough
reading of the manuscript and numerous comments and suggestions.

This work is based on observations obtained with the 6-m telescope of the
Special Astrophysical Observatory of the Russian Academy of Sciences. The
observations at the 6-metre BTA telescope were carried out with the financial
support of the Ministry of Education and Science of the Russian Federation
(agreement No. 14.619.21.0004, project ID RFMEFI61914X0004).

This work is based in part on observations made with the {\it Spitzer} Space
Telescope, which is operated by the Jet Propulsion Laboratory, California
Institute of Technology under a contract with NASA and on observations made
with the {\it Herschel}, which is an ESA space observatory with science
instruments provided by European-led Principal Investigator consortia and
with important participation from NASA.

This publication makes use of data products from the Two Micron All Sky
Survey, which is a joint project of the University of Massachusetts and the
Infrared Processing and Analysis Center/California Institute of Technology,
funded by the National Aeronautics and Space Administration and the National
Science Foundation and the National Science Foundation.

This research has made use of the SIMBAD database, operated at CDS,
Strasbourg, France.

The study of the ISM around vdB 130 was supported by the Russian
Foundation for Basic Research (projects 14-02-00027, 14-02-00472
and 15-07-04512) while the stellar population analysis of the
vdB~130 cluster was supported by RSCF grant No. 14-22-00041.

DSW was supported by the President of RF grant NSh-3620.2014.2 and by the OFN
RAS program `Interstellar and Intergalactic Medium: Active and Extended
Objects'.

AVM is also grateful for the financial support of the non-profit
`Dynasty' Foundation.

We used  the results of  H$\alpha$ interference observations performed with
the  125-cm reflector of the Crimean laboratory of Sternberg Astronomical
Institute of Moscow State University and the results of CO line radio
observations.

\label{lastpage}


\begin{thebibliography}{63}


\bibitem[\protect\citeauthoryear{Afanasiev \& Moiseev}{2005}]{scorpio}
Afanasiev V.L., Moiseev, A.V., 2005, Astron. Letters, 31, 194

\bibitem[\protect\citeauthoryear{Afanasiev \& Moiseev}{2011}]{scorpio2}
Afanasiev V.L., Moiseev A.V., 2011, Baltic Astronomy, 20, 363

\bibitem[\protect\citeauthoryear{Allen et al.}{2008}]{allen}
Allen M.G., Groves B.A., Dopita M.A., Sutherland R.S., Kewley L.J., 2008, ApJS, 178, 20

\bibitem[\protect\citeauthoryear{Anderson et~al.}{2010}]{and} Anderson L. D. et~al.,2010, A\&A, 518, 99

\bibitem[\protect\citeauthoryear{Arkhipova et~al.}{2013}]{ark13} Arkhipova V. P. et~al., 2013
 MNRAS, 432, 2273

\bibitem[\protect\citeauthoryear{Aniano et~al.}{2011}]{ani11} Aniano G.,
 Draine B.T., Gordon K.D., Sandstrom K., 2011, PASP, 123, 1218

\bibitem[\protect\citeauthoryear{Bendo et~al.}{2008}]{bendo} Bendo G. J. et~al., 2008, MNRAS, 389, 629

\bibitem[\protect\citeauthoryear{Blaha \& Humphreys}{1989}]{bla89} Blaha C., Humphreys R., 1989, AJ, 98, 1598

\bibitem[\protect\citeauthoryear{Bonatto et~al.}{2004}]{bbg04} Bonatto Ch., Bica E., Girardi L., 2004, A\&A, 415,
571

\bibitem[\protect\citeauthoryear{Cardelli et~al.}{1989}]{car89} Cardelli et al., 1989, ApJ, 345, 245.

\bibitem[\protect\citeauthoryear{Churchwell et~al.}{2006}]{Churchwell} Churchwell et al., 2006, ApJ, 649, 759.

\bibitem[\protect\citeauthoryear{Deharveng et~al.}{2009}]{deh} Deharveng L., Zavagno A., Schuller F., Caplan J., Pomare`s M., De
Breuck C., 2009, A\&A, 496, 177

\bibitem[\protect\citeauthoryear{Deharveng et~al.}{2010}]{bubblier}   Deharveng L. et~al, 2010, A\&A, 523, 6

\bibitem[\protect\citeauthoryear{Draine}{1978}]{Draine1978} Draine B. T., 1978, ApJS, 36, 595

\bibitem[\protect\citeauthoryear{Dutra et~al.}{2003}]{2Mext}
Dutra C.M., Santiago B.X., Bica E.L.D., Barbuy B. 2003,
MNRAS, V.338, P.253

\bibitem[\protect\citeauthoryear{Garmany \& Stencel}{1992}]{gar92} Garmany C.D.,Stencel C.D., 1992, A\&ASS, 94, 214

\bibitem[\protect\citeauthoryear{Girardi et~al.}{2002}]{padova}
Girardi L., Bertelli G., Bressan A., Chiosi C., Groenewegen M.A.T., Marigo
P., Salasnich, B., Weiss A. 2002, A\&A, 391, 195.

\bibitem[\protect\citeauthoryear{Gray \& Corbally}{2009}]{gray} Gray R.O., Corbally C.J., 2009, Stellar Spectral Classification, Princeton University Press, 565

\bibitem[\protect\citeauthoryear{Gutermuth et~al.}{2008}]{gut08} Gutermuth R.A. et~al, 2008, ApJ, 674, 356.

\bibitem[\protect\citeauthoryear{He et~al.}{1995}]{he95}He L.,
Whittet D.C.B., Kilkenny D., Spencer Jones J.H. 1995, ApJS, 101, 335.

\bibitem[\protect\citeauthoryear{Hog et al.}{2002}]{tycho2}
Hog E. et~al, 2000, A\&A, 355, 27.

\bibitem[\protect\citeauthoryear{Hora et~al.}{2011}]{cygX} Hora J.L., Smith H. A., Doering R. L., Spitzer Cygnus-X Survey Team, 2011, Bulletin of AAS,
43, AAS Meeting 217, 258.07

\bibitem[\protect\citeauthoryear{Hora  et~al.}{2009}]{hor09} Hora J.L. et al, 2009, Bulletin of AAS, 41, 498, AAS Meeting 213, 356.01

\bibitem[\protect\citeauthoryear{Kauffmann et~al.}{2003}]{Kauf03} Kauffmann G. et al., 2003, MNRAS, 346, 1055

\bibitem[\protect\citeauthoryear{Kewley et~al.}{2001}]{Kewley2001} Kewley L.J., Dopita M.A., Sutherland R.S., Heisler C.A., Trevena J.,
2001, ApJ, 556, 121

\bibitem[\protect\citeauthoryear{Kewley et~al.}{2006}]{Kewley2006} Kewley L.J., Groves B.A., Kauffmann G., Heckman T., 2006, MNRAS, 372, 961

\bibitem[\protect\citeauthoryear{Kharchenko et~al.}{2013}]{khar13} Kharchenko N.V.,
Piskunov A.E., Schilbach E., Roser S., Scholz R.-D., 2013, A\&A, 558, A53.

\bibitem[\protect\citeauthoryear{King}{1962}]{king62} King I.R., 1962,
AJ, 67, 471

\bibitem[\protect\citeauthoryear{Le Duigou \& Kn\"{o}dlseder}{2002}]{Led02}  Le Duigou J.-M., Kn\"{o}dlseder J., 2002, A\&A, 392, 869

\bibitem[\protect\citeauthoryear{Lozinskaya \& Sitnik}{1988}]{loz88}  Lozinskaya T.A., Sitnik T.G., 1988, SvAL, 14, 100

\bibitem[\protect\citeauthoryear{Lozinskaya \& Repin}{1990}]{loz90} Lozinskaya T.A., Repin S.V., 1990, SvA, 34, 580

\bibitem[\protect\citeauthoryear{Lozinskaya et~al.}{1997}]{loz97} Lozinskaya T. A., Pravdikova V. V., Sitnik T. G., Esipov V. F.,
Mel'nikov V. V., 1997, Astron. Lett., 23, 450.

\bibitem[\protect\citeauthoryear{Lozinskaya et~al.}{1998}]{loz98} Lozinskaya T. A., Pravdikova V. V., Sitnik T. G., Esipov, V.
F., Mel'nikov V. V., 1998, Astron. Reports, 42..453.

\bibitem[\protect\citeauthoryear{Magakian}{2003}]{mag} Magakian T.Y., 2003, A\&A, 399, 141.

\bibitem[\protect\citeauthoryear{Martin \& Whittet}{1990}]{mar90} Martin, Whittet, 1990, ApJ, 357,
113.

\bibitem[\protect\citeauthoryear{Melnik \& Dambis}{2009}]{MD2009} Melnik A.M., Dambis A.K.,
2009, MNRAS, 400, 518

\bibitem[\protect\citeauthoryear{Moffat \& Schmidt-Kaler}{1976}] {moffat} Moffat A.F.J.,
Schmidt-Kaler T., 1976, A\&A, 48, 115

\bibitem[\protect\citeauthoryear{Molinari et al.}{2010}] {Molinari10} Molinari et al., 2010, PASP, 122, 314

\bibitem[\protect\citeauthoryear{Motte et al.}{2007}] {Motte2007} Motte F., Bontemps S., Schilke P., Schneider N.,
Menten K. M., Brogui\`ere D., 2007, A\&A, 476, 1243

\bibitem[\protect\citeauthoryear{Pavlyuchenkov  et~al.}{2013}]{kirsanova} Pavlyuchenkov Ya. N.,
Kirsanova M. S., Wiebe D. S. 2013, ARep, 57, 573

\bibitem[\protect\citeauthoryear{Pilyugin \& Mattsson}{2011}]{abund}  Pilyugin L.S., Mattsson L., 2011, MNRAS, 412, 1145

\bibitem[\protect\citeauthoryear{Racine}{1968}]{rac68}  Racine R., 1968, AJ, 73, 233

\bibitem[\protect\citeauthoryear{Racine}{1974}]{rac74}  Racine R., 1974, AJ, 79, 945

\bibitem[\protect\citeauthoryear{Roeser et al.}{2010}]{ppmxl} Roeser S., Demleitner M., Schilbach
E., 2010, AJ., 139, 2440

\bibitem[\protect\citeauthoryear{Saken et~al.} {1992}]{sak92}
Saken J.M., Shull J.M., Garmany C.D., Nichols-Bohlin J., Fesen R., 1992, ApJ,
397, 537

\bibitem[\protect\citeauthoryear{Schlafly \& Finkbeiner}{2011}]{sf2011} Schlafly E.F., Finkbeiner D.P., 2011, ApJ 737,
id.103

\bibitem[\protect\citeauthoryear{Schlegel et al.}{1998}]{sfd1998} Schlegel D.J., Finkbeiner D.P., Davis M., 1998, ApJ 500,
525

\bibitem[\protect\citeauthoryear{Schneider et~al.}{2006}] {sch06} Schneider, N., Bontemps, S., Simon, R., Jakob, H., Motte, F., Miller, M., Kramer, C., Stutzki, J.,
2006, A\&A, 458, 855

\bibitem[\protect\citeauthoryear{Schneider et~al.}{2007}]{sch07} Schneider N., Simon R., Bontemps S., Comeron F., Motte F., 2007,
A\&A, 474, 873

\bibitem[\protect\citeauthoryear{Schneider et~al.}{2011}] {sch11} Schneider N. et~al, 2011, A\&A, 529, 1

\bibitem[\protect\citeauthoryear{Skrutskie et~al.}{2006}] {skr06} Skrutskie M.F. et~al, 2006, AJ, 131,
1163

\bibitem[\protect\citeauthoryear{Silva \& Cornell}{1992}]{sil92} Silva D.R., Cornell M.E.,1992, ApJSS 81, 865

\bibitem[\protect\citeauthoryear{Sitnik}{2003}]{sit03} Sitnik T.G., 2003, Astron. Lett., 29, 311

\bibitem[\protect\citeauthoryear{Sitnik \& Mel'nik}{1996}]{sit96} Sitnik T.G., Mel'nik A.M., 1996, Astron. Lett., 22, 122

\bibitem[\protect\citeauthoryear{Valdes et~al.}{2004}]{val04} Valdes F., Gupta R., Rose J.A., Singh, H.P., Bell, D.J., 2004, A\&ASS, 152, 251

\bibitem[\protect\citeauthoryear{van den Bergh}{1966}]{van66}  van den Bergh S., 1966, AJ, 71, 990

\bibitem[\protect\citeauthoryear{Watson et~al.}{2008}]{bubble} Watson C. et~al, 2008, ApJ, 681, 1341

\bibitem[\protect\citeauthoryear{Ybarra \& Lada}{2009}]{ybarra2009} Ybarra J. E., Lada E. A., 2009, ApJL, 695, L120

\bibitem[\protect\citeauthoryear{Ybarra et~al.}{2014}]{ybarra2014} Ybarra J. E., Tapia, M., Roman-Zuniga, C.,
Lada, E., 2014, ApJL, 794, 25

\bibitem[\protect\citeauthoryear{Zabolotskikh et~al.}{2002}]{zabol02} Zabolotskikh M.V.,
Rastorguev A.S., Dambis A.K., 2002, Astron. Lett., 28, 454

\bibitem[\protect\citeauthoryear{Zacharias et~al.}{2013}]{ucac4} Zacharias N.,
 Finch C.T., Girard T.M., Henden A., Bartlett J.L., Monet D.G. Zacharias M.I.,
 2013, AJ, 145, 44


\end{thebibliography}
\end{document}